\documentclass[letterpaper,english,reprint,superscriptaddress,longbibliography]{revtex4-1}
\usepackage{lmodern}

\usepackage[T1]{fontenc}
\usepackage[utf8]{inputenc}
\usepackage{color}
\usepackage{babel}
\usepackage{amsmath}
\usepackage{amssymb}
\usepackage{graphicx}
\usepackage[unicode=true,pdfusetitle,
 bookmarks=false,
 breaklinks=false,pdfborder={0 0 0},pdfborderstyle={},backref=false,colorlinks=false]
 {hyperref}

\makeatletter

\pdfpageheight\paperheight
\pdfpagewidth\paperwidth

\makeatother

\begin{document}
\title{Resonant two-laser spin-state spectroscopy of a negatively charged
quantum dot-microcavity system with a cold permanent magnet}
\author{P. Steindl}
\email{steindl@physics.leidenuniv.nl}

\affiliation{Huygens-Kamerlingh Onnes Laboratory, Leiden University, P.O. Box 9504,
2300 RA Leiden, The Netherlands}
\author{T. van der Ent}
\affiliation{Huygens-Kamerlingh Onnes Laboratory, Leiden University, P.O. Box 9504,
2300 RA Leiden, The Netherlands}
\author{H. van der Meer}
\affiliation{Huygens-Kamerlingh Onnes Laboratory, Leiden University, P.O. Box 9504,
2300 RA Leiden, The Netherlands}
\author{J.A. Frey}
\affiliation{Department of Physics, University of California, Santa Barbara, California
93106, USA}
\author{J. Norman}
\affiliation{Department of Electrical \& Computer Engineering, University of California,
Santa Barbara, California 93106, USA}
\author{J.E. Bowers}
\affiliation{Department of Electrical \& Computer Engineering, University of California,
Santa Barbara, California 93106, USA}
\author{D. Bouwmeester}
\affiliation{Huygens-Kamerlingh Onnes Laboratory, Leiden University, P.O. Box 9504,
2300 RA Leiden, The Netherlands}
\affiliation{Department of Physics, University of California, Santa Barbara, California
93106, USA}
\author{W. Löffler}
\email{loeffler@physics.leidenuniv.nl}

\affiliation{Huygens-Kamerlingh Onnes Laboratory, Leiden University, P.O. Box 9504,
2300 RA Leiden, The Netherlands}
\begin{abstract}
A high-efficiency spin-photon interface is an essential piece of quantum
hardware necessary for various quantum technologies. Self-assembled
InGaAs quantum dots have excellent optical properties, if embedded
into an optical micro-cavity they can show near-deterministic spin-photon
entanglement and spin readout, but an external magnetic field is required
to address the individual spin states, which usually is done using
a superconducting magnet. Here, we show a compact cryogenically compatible
SmCo magnet design that delivers $475\,\mathrm{\textrm{mT}}$ in-plane
Voigt geometry magnetic field at $5\,\mathrm{\textrm{K}}$, which
is suitable to lift the energy degeneracy of the electron spin states
and trion transitions of a single InGaAs quantum dot. This quantum
dot is embedded in a birefringent high-finesse optical micro-cavity
which enables efficient collection of single photons emitted by the
quantum dot. We demonstrate spin-state manipulation by addressing
the trion transitions with a single and two laser fields. The experimental
data agrees well to our model which covers single- and two-laser cross-polarized
resonance fluorescence, Purcell enhancement in a birefringent cavity,
and variation of the laser powers.
\end{abstract}
\maketitle

\section{introduction}

An efficient, tunable spin-photon interface that allows high fidelity
entanglement of spin qubits with flying qubits, photons, lies at the
heart of many building blocks of distributed quantum technologies
\cite{Awschalom2018}, ranging from quantum repeaters \cite{Kimble2008},
photonic gates \cite{Koshino2010,Rosenblum2011}, to the generation
of photonic cluster states \cite{Lindner2009,Schwartz2016,Coste2022}.
Further, to secure connectivity within the quantum network, an ideal
spin-photon interface requires near-unity collection efficiency, therefore
an atom or semiconductor quantum dot (QD) carrying a single spin as
a quantum memory is integrated into photonic structures such as optical
microcavities cavities, where recently $57\,\%$ in-fiber photon collection
efficiency has been achieved \cite{Tomm2021}.

Within the pool of promising systems, singly-charged excitonic complexes
of optically active QD devices in III-V materials \cite{Warburton2013}
combines near-unity quantum efficiency, excellent zero-phonon line
emission at cryogenic temperatures \cite{Favero2003} with nearly
lifetime-limited optical linewidth \cite{Kuhlmann2015}. This, in
combination with sub-nanosecond Purcell-enhanced lifetimes, enabled
GHz-scale generation rates of indistinguishable single-photons \cite{Santori2002,He2013,Ding2016,Somaschi2016,Senellart2017,Hilaire2020,Tomm2021},
robust polarization selection rules \cite{Bayer2002,Stockill2017},
and simple on-chip integration facilitating stable-long term operation
and tuneability.

The singly-charged quantum dot can be optically excited to the trion
state, if this is done with linearly polarized light, the spin state
of the resident electron is transferred to the trion hole spin by
the optical selection rules. If the trion decays, it will emit a single
circularly polarized photon with a helicity depending on the hole
spin state, Fig. \ref{Fig1:Magnet_rendering}(a). To achieve selective
spin addressability which is necessary for spin initialization and
readout, the QD is typically placed in an external in-plane (Voigt
geometry) magnetic field \cite{Emary2007,Xu2007}, which induces Zeeman
splitting of the spin states and trion transitions \cite{Bayer2002}.
The magnetic field modifies the eigenstates of the system and the
optical selection rules, and four optical transitions are possible
(see Fig. \ref{Fig1:Magnet_rendering}(b)), which are now linearly
polarized. The electron and trion spin, as well as the photon polarization,
are now connected by the modified optical selection rules. We obtain
two intertwined $\Lambda$ systems which can be used with steady-state
light fields for spin initialization \cite{Xu2007,Emary2007}, arbitrary
spin ground state superposition generation \cite{Xu2008}, or dynamical
spin decoupling from the nuclear bath \cite{Xu2009}. 

This spin manipulation gets harder if the quantum dot is placed in
a non-polarization degenerate (birefringent) micro-cavity \cite{Tomm2021,Steindl2021_PRL,Hilaire2020}.
Here we show two-laser resonant spectroscopy \cite{Xu2007,Kroner2008PRB}
of a single spin in a single quantum dot in such a birefringent cavity,
and use cross-polarized collection of single photons. We use a simple
“set-and-forget” cryogenic permanent magnet assembly to apply the
magnetic field, and we are able to derive the spin dynamics by comparison
to a theoretical model.

\section{Permanent magnet assembly}

\begin{figure}
\begin{centering}
\includegraphics[width=1\columnwidth]{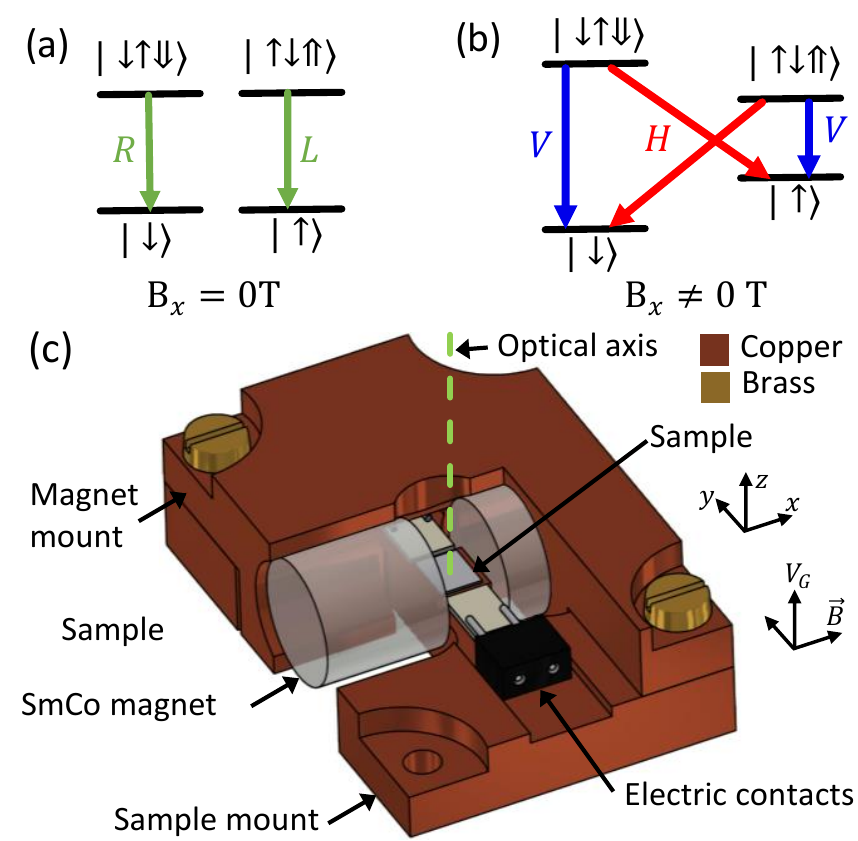}
\par\end{centering}
\caption{Energy level schemes of negatively charged QD and magnetic assembly.
Optical selection rules of trion transitions without (a) and with
(b) an external in-plane magnetic field. (c) Schematic of the permanent
magnet assembly. \label{Fig1:Magnet_rendering}}
\end{figure}
Magneto-optical quantum dot-based experiments usually rely on large
and complex superconducting magnets \cite{Coste2022,Huber2019}, which
generate strong magnetic fields but require both a stabilized current
source and cryogenic temperatures. However, many experiments need
only a ``set-and-forget'' static magnetic field of around $500\,\mathrm{mT},$
which can be achieved with compact strong permanent magnets cooled
down together with the quantum dot device \cite{Androvitsaneas2016,Androvitsaneas2019,Adambukulam2021}.
Unfortunately, many rare-earth magnetic materials such as NdFeB \cite{Fredricksen2003}
suffer at cryogenic temperatures from spin reorientation \cite{Tokuhara1985}
which lowers the effective magnetic field \cite{He2013_magnet} and
tilts the easy axis of the magnetic assembly \cite{Hara2004,Fredricksen2003}.
Especially, losing control over the magnetic field direction is problematic
with quantum dots since it affects mixing between dark and bright
states and thus changes both transition energies and optical selection
rules \cite{Bayer2000}.

To build our permanent magnet assembly, we have chosen from the strongest
commercially available magnetic materials \cite{Miyake2018,Wang2021}
SmCo (grade 2:17) magnets with a room temperature remanence of $1.03\,\mathrm{T}$.
This industrially used magnetic system is known for its high Curie
temperature (over $800\,^{\circ}\mathrm{C}$) and high magnetocrystalline
anisotropy \cite{Zhou2000,Gutfleisch2000} excellent for high-temperature
applications in several fields \cite{Gutfleisch2011,Duerrschnabel2017,Guo2022}.
Especially it is used above the Curie temperatures of NdFeB of $310\,^{\circ}\mathrm{C}$
\cite{Gutfleisch2000}, where current NdFeB-based magnets have relatively
poor intrinsic magnetic properties. Moreover, due to low temperature-dependence
of remanence and coercivity \cite{Durst1986,Liu2019,Wang2021}, SmCo-based
magnets also show excellent thermal stability of the remanence with
near-linear dependence \cite{Durst1986,He2013_magnet} down to $4.2\,\mathrm{K}$.
This is in contrast to other common rare-earth magnet compounds such
as NdFeB \cite{Fredricksen2003}, where the remanence at temperatures
below $135\,\mathrm{K}$, depending on the specific material composition
\cite{He2013_magnet}, decreases rapidly by several percent due to
the spin-reorientation transition \cite{Tokuhara1985}. 

Our permanent magnet assembly in Fig. \ref{Fig1:Magnet_rendering}(c)
is designed to fit on top of a $XYZ$ piezo motor assembly in a standard
closed-cycle cryostat with optical access via an ambient-temperature
long working distance objective, which restricts its physical dimensions
to approximately $1\,\mathrm{cm}$ in height. Thus, we built the assembly
from two $9\times9\,\mathrm{mm}$ commercially available rod-shaped
SmCo magnets separated by a $4.5\,\mathrm{\textrm{mm}}$ air gap embedded
in a $36\times24\times10.8\,\mathrm{mm}$ copper housing. Due to the
large remanence ($1.03\,\mathrm{T}$) and small air gap, the assembly
in the center of the gap produces a homogeneous magnetic field of
about $500\,\mathrm{\textrm{mT}}$, as discussed in Appendix \ref{sec:Room-temperature-permanent}.
The assembly is rigidly attached by brass screws to the H-shaped copper
sample mount, where the quantum dot device is horizontally placed
in the center of the air gap such that the magnetic field is in-plane
(Voigt geometry). The assembly contains electrical contacts to apply
a bias voltage $V_{\textrm{G}}$ to the device. It has a low weight
of $69\,\mathrm{\textrm{g}}$ (including $4.8\,\mathrm{g}$ per magnet),
compatible with standard nanopositioners allowing for fine-tuning
of the sample position with respect to the optical axis. 

The magnetic mount is then cooled down together with the sample to
approximately $5\,\mathrm{K}.$ Since in SmCo, the spin reorientation
transition was reported to be stable down to $10\,\mathrm{K}$ \cite{He2013_magnet},
we do not expect magnetization axis changes and assume only a small
magnetic field drop of $5\,\%$ between the room and cryogenic temperatures
\cite{Fredricksen2003}. This makes SmCo an ideal material choice
for strong homogeneous cryogenic magnets, in our case delivering about
$475\,\mathrm{mT}$ at $5\,\mathrm{K}$.

\section{Spin-state determination}

\begin{figure}
\begin{centering}
\includegraphics[width=1\columnwidth]{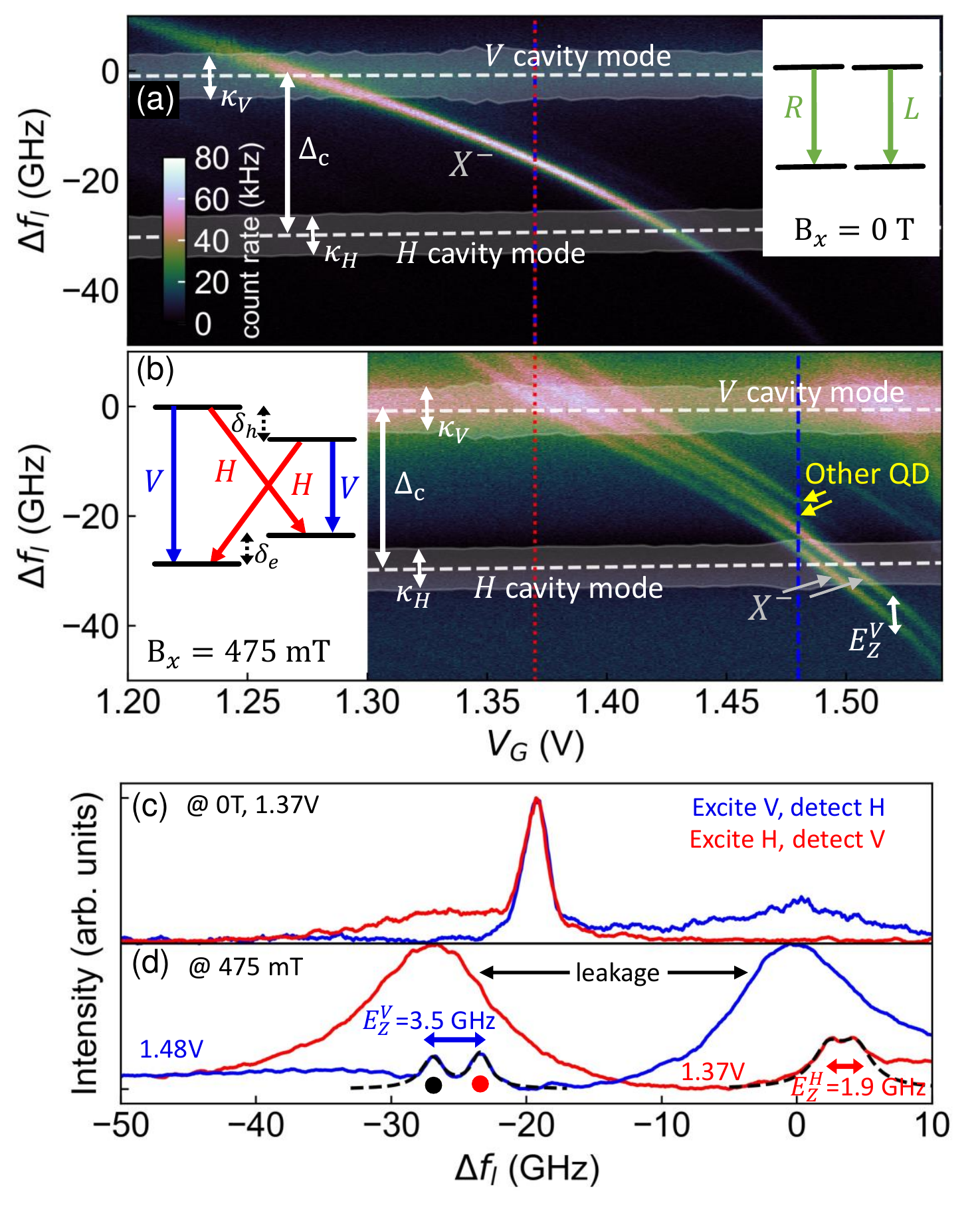}
\par\end{centering}
\caption{Resonant reflection as a function of laser frequency and gate voltage
without (a) and with (b) an in-plane external magnetic field, plotted
with the same color scale. The excitation laser is polarized along
the $V$ cavity axis and reflected laser light is filtered out using
a crossed $H$ polarizer, to select photons emitted by the QD trion.
Insets show the corresponding optical selection rules. Dashed lines
indicate the cavity resonance frequencies with the cavity decay rates
$\kappa_{V}$, $\kappa_{H}$, determined using semi-classical model
fits \cite{Snijders2020}. Panels (c, d) show cross-sectional plots
without and with magnetic field for two excitation polarizations (blue:
excitation along $V$ cavity mode, red: $H$) at voltages $1.37\,\mathrm{V}$
or $1.48\,\mathrm{V}$, indicated by the vertical lines in panels
(a, b). The Zeeman splittings determined from Lorentzian fits (black
dashed lines) are given. The excitation power in front of excitation
objective is $2\,\mathrm{nW}$, laser scanning speed $41\,\mathrm{GHz}\mathrm{/s}$.\label{Fig2:Voltage_scan_0T_vs_500mT}}
\end{figure}
We study self-assembled InGaAs quantum dots emitting around $\lambda=935.5\,\mathrm{nm}$,
embedded in $\sim\lambda$ thick GaAs planar cavity, surrounded by
two distributed Bragg reflectors (DBR): 26 pairs of $\lambda/4$ thick
GaAs/$\text{A\ensuremath{\text{l}_{0.90}}G\ensuremath{\text{a}_{0.10}}As}$
layers from the top and 13 pairs of GaAs/AlAs layers and 16 pairs
of GaAs/$\text{A\ensuremath{\text{l}_{0.90}}G\ensuremath{\text{a}_{0.10}}As}$
layers at the bottom \cite{Snijders2018,Steindl2021_PRL}. The single
QD layer is embedded in a\textit{ p-i-n} junction, separated from
the electron reservoir by a $31.8\,\mathrm{nm}$ thick tunnel barrier
including a $21.8\,\mathrm{nm}$ thick $\text{A\ensuremath{\text{l}_{0.45}}G\ensuremath{\text{a}_{0.55}}As}$
electron blocking layer designed to allow single electron charging
of the QD \cite{Heiss2008,Heiss2010}. A voltage bias $V_{\textrm{G}}$
applied over the diode allows for charge-control of the ground state
of the quantum dot and also to fine-tune the QD transition energies
into resonance with the optical cavity mode. The optical in-plane
cavity mode confinement is achieved by an oxide aperture, we fabricate
216 cavities per device \cite{Frey_phdthesis} and select a suitable
one with (i) a quantum dot well-coupled to the cavity mode and (ii)
low birefringence of the fundamental mode, for the device studied
here, the two linearly-polarized modes cavity modes ($H$ and $V$
modes) are split by $\Delta_{\textrm{c}}=28\,\mathrm{GHz}$.

First, we cool down the device to 5 K without the SmCo magnet assembly
in a closed-cycle cryostat. For resonant laser spectroscopy, we use
a cross-polarization laser extinction method with laser rejection
better than $10^{6}$ \cite{Steindl2023_PER}. Using a free-space
polarizer and half-waveplate, the polarization of the excitation laser
is aligned along the $V$ cavity polarization axis, and the light
reflected from the cavity is recorded with a single-photon detector
after passing again the half-wave plate and the crossed polarizer.
In Fig. \ref{Fig2:Voltage_scan_0T_vs_500mT}(a), we show a fluorescence
map of this device measured in the cross-polarization scheme as a
function of the laser frequency detuning from the $V$-polarized cavity
mode resonance $\Delta f_{l}$ and applied bias voltage $V_{\textrm{G}}$.
We observe a single emission line which is shifted by the quantum-confined
Stark effect, the line is in resonance with the $V$ cavity mode at
around $1.25\,\mathrm{V}$ and with the $H$ cavity mode at around
$1.40\,\mathrm{V}$. The same line is visible also if the excitation
and detection polarization are swapped, see the cross-sectional plot
in Fig. \ref{Fig2:Voltage_scan_0T_vs_500mT}(c). The fact that we
observe the same single line under both perpendicular polarizations
and that it is coupled to both fundamental cavity modes, suggests
that the emitted photons are circularly polarized and originate from
the charged exciton $X^{-}$.

Now we cool down the device with the SmCo magnet assembly, to lift
the energy degeneracy of the trion transitions. In this scenario,
with the energy level scheme in Fig. \ref{Fig2:Voltage_scan_0T_vs_500mT}(b),
the optical selection rules are modified by the in-plane magnetic
field from circular to linear polarization. Thus the scanning excitation
laser polarized along the $V$ cavity mode can only resonantly address
$V$-polarized transitions, i.e., $\left|\downarrow\right\rangle \rightarrow\left|\downarrow\uparrow\Downarrow\right\rangle $
and $\left|\uparrow\right\rangle \rightarrow\left|\uparrow\downarrow\Uparrow\right\rangle $,
therefore we expect to observe a pair of lines Zeeman-split by the
energy $E_{\textrm{Z}}^{V}=\delta_{\textrm{e}}+\delta_{\textrm{h}}$.
Without cavity enhancement, each of the excited trion states radiatively
decays with equal probability (by cavity Purcell enhancement, however,
this is modified) into the single-spin ground state by emission of
a single photon with either $V$ or $H$ polarization depending on
the excited and ground states, as depicted in Fig. \ref{Fig2:Voltage_scan_0T_vs_500mT}(b).
Because we measure in cross-polarization, we filter out the emitted
$V$-polarized single photons and detect only photons emitted by the
$\left|\downarrow\uparrow\Downarrow\right\rangle \rightarrow\left|\uparrow\right\rangle $
and $\left|\uparrow\downarrow\Uparrow\right\rangle \rightarrow\left|\downarrow\right\rangle $
transitions. Thus, the total detected rate is reduced to half of that
without magnetic field. Similarly, the scanning laser polarized along
the $H$ cavity mode excites only $\left|\uparrow\right\rangle \rightarrow\left|\downarrow\uparrow\Downarrow\right\rangle $
and $\left|\downarrow\right\rangle \rightarrow\left|\uparrow\downarrow\Uparrow\right\rangle $,
and we observe again a pair of fluorescence lines, this time Zeeman
split by $E_{\textrm{Z}}^{H}=|\delta_{\textrm{e}}-\delta_{\textrm{h}}|$.
Note that in Fig. \ref{Fig2:Voltage_scan_0T_vs_500mT}(b) we observe
two pairs of emission lines which originate from two different QDs.
We focus only on the brighter QD, corresponding to the clear transition
in Fig. \ref{Fig2:Voltage_scan_0T_vs_500mT}(a). In agreement with
the trion energy level scheme, the trion transitions exhibit a different
Zeeman splitting of $E_{\textrm{Z}}^{V}=3.5\pm0.1\,\mathrm{GH\mathrm{z}}$
under $V$- and $E_{\textrm{Z}}^{H}=1.9\pm0.1\,\mathrm{GHz}$ $H$-polarization
excitation. This Zeeman splitting was extracted by Lorentizan fits
to laser frequency scans shown in Fig. \ref{Fig2:Voltage_scan_0T_vs_500mT}(d),
which allows us to estimate \cite{Bennett2013} the electron and hole
g-factors, we obtain $|g_{\textrm{e}}|=0.39$ and $|g_{\textrm{h}}|=0.12$;
these values agree to literature values for small InGaAs QDs \cite{Nakaoka2004}.
We also observe a $25\,\textrm{GHz}$ energy average shift of the
QD emission caused by a combination of the diamagnetic shift (around
$0.5\,\textrm{GHz}$ assuming diamagnetic constant $-9.4\,\textrm{\ensuremath{\mu}eV/\ensuremath{T^{2}}}$
\cite{Kroner2008PRB}), and temperature/strain induced band-gap changes
between consecutive cooldowns. Note that we also observe a broad emission,
which is most likely due to cavity-enhanced fluorescence in combination
with imperfect polarization alignment and due to our limited cross-polarization
extinction ratio of $4\times10^{6}$ \cite{Steindl2023_PER}. 

\section{Two-color resonant laser excitation}

\begin{figure}
\begin{centering}
\includegraphics[width=1\columnwidth]{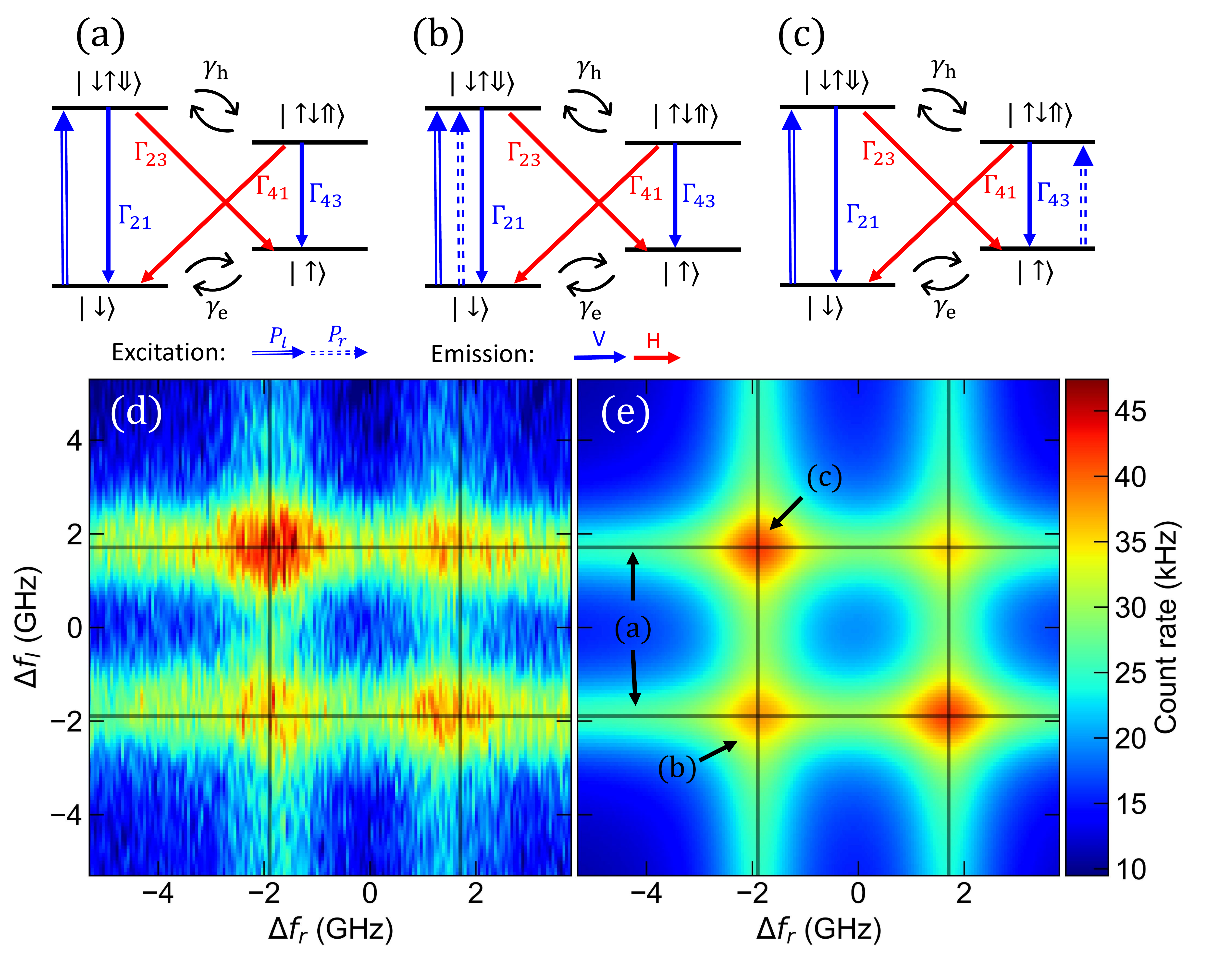}
\par\end{centering}
\caption{Two-color laser trion spectroscopy with a magnetic field in Voigt
geometry. (a-c) Comparison of three different trion excitation conditions:
single-laser excitation (a), two same-frequency lasers (b), and (c)
each laser addresses different trion transitions. Experiment (d) and
model (e) data for of two-color experiments, for $P_{l}=2.1\:\mathrm{nW}$
pump laser power and $P_{r}=2.0\,\mathrm{nW}$ repump laser power,
the black lines indicate the QD trion transition frequencies. \label{Fig3:2LRF_scan_theory_vs_exp}}
\end{figure}

Now, we demonstrate spin-state manipulation using two individually
tunable narrow-linewidth lasers. For a high-degree cross-polarization
extinction ratio, we perform resonance fluorescence spectroscopy in
the vicinity of the $H$-cavity mode ($V_{\textrm{G}}=1.49\,\mathrm{V}$,
i.e., we focus on the transitions marked by dots in Fig. \ref{Fig2:Voltage_scan_0T_vs_500mT}(d)),
and we use $V$ polarization of both excitation lasers. In Fig. \ref{Fig3:2LRF_scan_theory_vs_exp}(d),
we show a reflection map measured in cross-polarization as a function
of both laser frequencies $f_{l}$ (pump) and $f_{r}$ (repump). The
horizontal and vertical lines indicate the trion transition frequencies.
Where these frequencies intersect interesting dynamics occurs. First,
the nodes oriented along the diagonal represent a condition where
both lasers are resonant with the same transition corresponding to
the excitation scheme depicted in Fig. \ref{Fig3:2LRF_scan_theory_vs_exp}(b).
We will call this configuration two-laser resonant excitation (2LRE).
The system dynamics under this excitation is equivalent to single-laser
excitation (1LRE) with stronger emission due to the higher driving
power of $P_{l}+P_{r}$. The anti-diagonally oriented nodes correspond
to emission under two-color excitation where each laser pumps a distinct
transition {[}Fig. \ref{Fig3:2LRF_scan_theory_vs_exp}(c){]}; we refer
to this scheme as two-color resonant excitation (2CRE) \cite{Kroner2008PRB}.
For clarity, we further focus only on the situation where the first
laser of constant power $P_{l}$ continuously pumps the $\left|\downarrow\right\rangle \rightarrow\left|\downarrow\uparrow\Downarrow\right\rangle $
transition. Due to cross-polarization detection, we observe only $H$-polarized
emission from the $\left|\downarrow\uparrow\Downarrow\right\rangle \rightarrow\left|\uparrow\right\rangle $
transition, a signature of population shelving into the $\left|\downarrow\right\rangle $
spin state. This shelved population is repumped, and thus, the total
(detected) single-photon rate increased by re-pumping the $\left|\uparrow\right\rangle \rightarrow\left|\uparrow\downarrow\Uparrow\right\rangle $
transition with the second laser, and we observe a higher photon rate
at the anti-diagonal nodes in Fig. \ref{Fig3:2LRF_scan_theory_vs_exp}(d).

To gain a more precise knowledge of the magnitude of the spontaneous
decay rates $\Gamma_{xy}$ as well as electron and hole spin-flip
rates $\gamma_{\textrm{e}}$ and $\gamma_{\textrm{h}}$ involved in
the system dynamics, we compare our experiments to a model which is
derived in the Appendix \ref{subsec:Semi-classical-model-of}. For
a laser power below the saturation power $P_{\textrm{c}}$, the model
is derived from the rate equations describing the steady-state two-scanning
lasers pump of the trion energy scheme in Fig. \ref{Fig3:2LRF_scan_theory_vs_exp}.
The trion transitions are modeled as two coupled $\Lambda$ systems
with asymmetric $V$ and $H$-polarized radiative transition rates
due to cavity enhancement of the latter. A careful analysis of the
model parameters and comparison to our experimental results allows
us to determine the electron spin-flip rates to be $\gamma_{\textrm{e}}\approx2.5\,\mathrm{MHz}$,
while the hole spin-flip rate cannot be determined because of the
short lifetime of the excited trion states, as expected. Further we
obtain lifetimes of $\Gamma_{21}=2.1\,\textrm{GHz},\Gamma_{43}=2.7\mathrm{\,GHz}$,
and $\Gamma_{23}=\Gamma_{41}=0.8\,\mathrm{GHz}$. Similar spin-flip
rates were reported in earlier resonant two-color trion spectroscopy
without cavity \cite{Kroner2008PRB}; the cavity-enhanced radiative
rates $\Gamma_{21}$, $\Gamma_{43}$ agree with our power-broadening
analysis, see Appendix \textcolor{black}{\ref{subsec:Single-laser-resonant}}. 

\begin{figure}
\begin{centering}
\includegraphics[width=1\columnwidth]{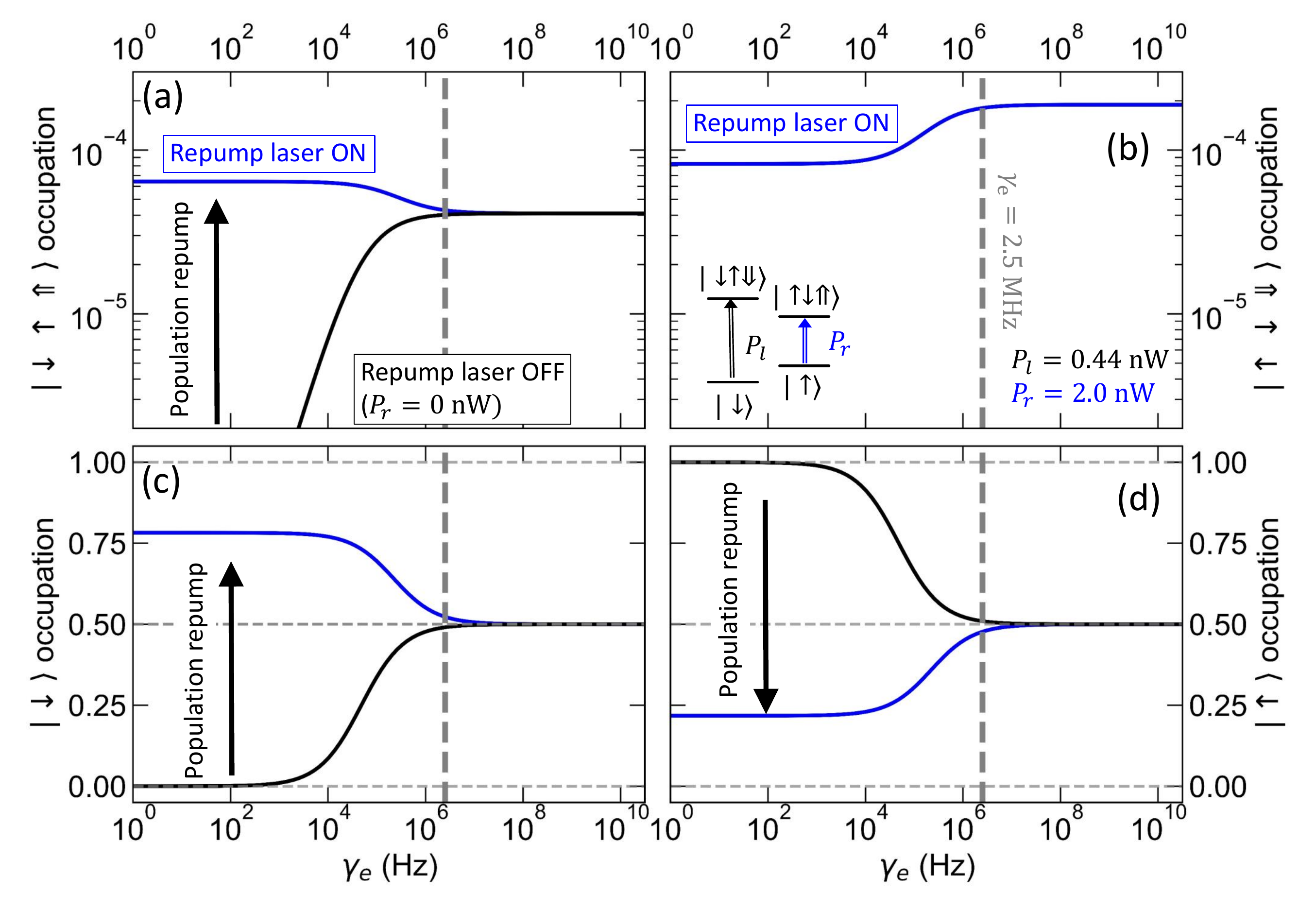}
\par\end{centering}
\caption{Steady-state trion states (a, b) and electron ground-state spin (c,
d) occupation probability as a function of electron spin-flip rate,
with (blue) and without (blue) repump laser. The dashed lines show
the determined spin-flip rate of $\gamma_{\textrm{e}}\approx2.5\,\mathrm{MHz}$.\label{Fig4:Model_occupation_vs_spinflips}}
\end{figure}

Figure \ref{Fig4:Model_occupation_vs_spinflips} shows spin-flip rate
dependency of the steady state occupation of the trion and electron
spin states predicted by our theory. In the simulation with varied
$\gamma_{\textrm{e}}$, we used system parameters found above together
with laser powers $P_{l}=0.44\,\textrm{nW}$ and $P_{r}=2.0\,\textrm{nW}$
to demonstrate spin pumping. First, if the electron spin-flip rate
is small (below $1\,\textrm{kHz}$), the weak pump laser light initializes
the spin state $\left|\uparrow\right\rangle $. By optical repumping
with the second laser on resonance with $\left|\uparrow\right\rangle \rightarrow\left|\uparrow\downarrow\Uparrow\right\rangle $,
the shelved spin population can be largely transferred from $\left|\uparrow\right\rangle $
into $\left|\downarrow\right\rangle $ as demonstrated in Fig. \ref{Fig4:Model_occupation_vs_spinflips}(c,d).
Due to the optical repumping, the resonant absorption on spin $\left|\uparrow\right\rangle $
becomes again possible, leading experimentally in the recovery of
transmission signal at the resonant frequency with $\left|\downarrow\right\rangle \rightarrow\left|\downarrow\uparrow\Downarrow\right\rangle $
\cite{Kroner2008PRB,Atature2006}. Our simulation for the determined
spin-flip rate of $\gamma_{\textrm{e}}\approx2.5\,\mathrm{MHz}$ shows
that the electron spin-flip leads to a comparable spin population
of both ground states even without repumping laser field, making conclusive
absorption measurements difficult because of the small change between
ground state populations with and without optical repumping. However,
the spin repumping from $\left|\uparrow\right\rangle $ is accompanied
by the population of $\left|\uparrow\downarrow\Uparrow\right\rangle $
resulting in extra emission from this spin state. Importantly, the
presence of this extra emission is independent of the ground state
spin-flip rate and can be thus used as a signature of optical spin
repumping. Moreover, at low $\gamma_{\textrm{e}}$, the emission following
the spin repumping benefits also from the extra excited state population
of the state $\left|\downarrow\uparrow\Downarrow\right\rangle $,
see Fig. \ref{Fig4:Model_occupation_vs_spinflips}(a).

\begin{figure}
\begin{centering}
\includegraphics[width=1\columnwidth]{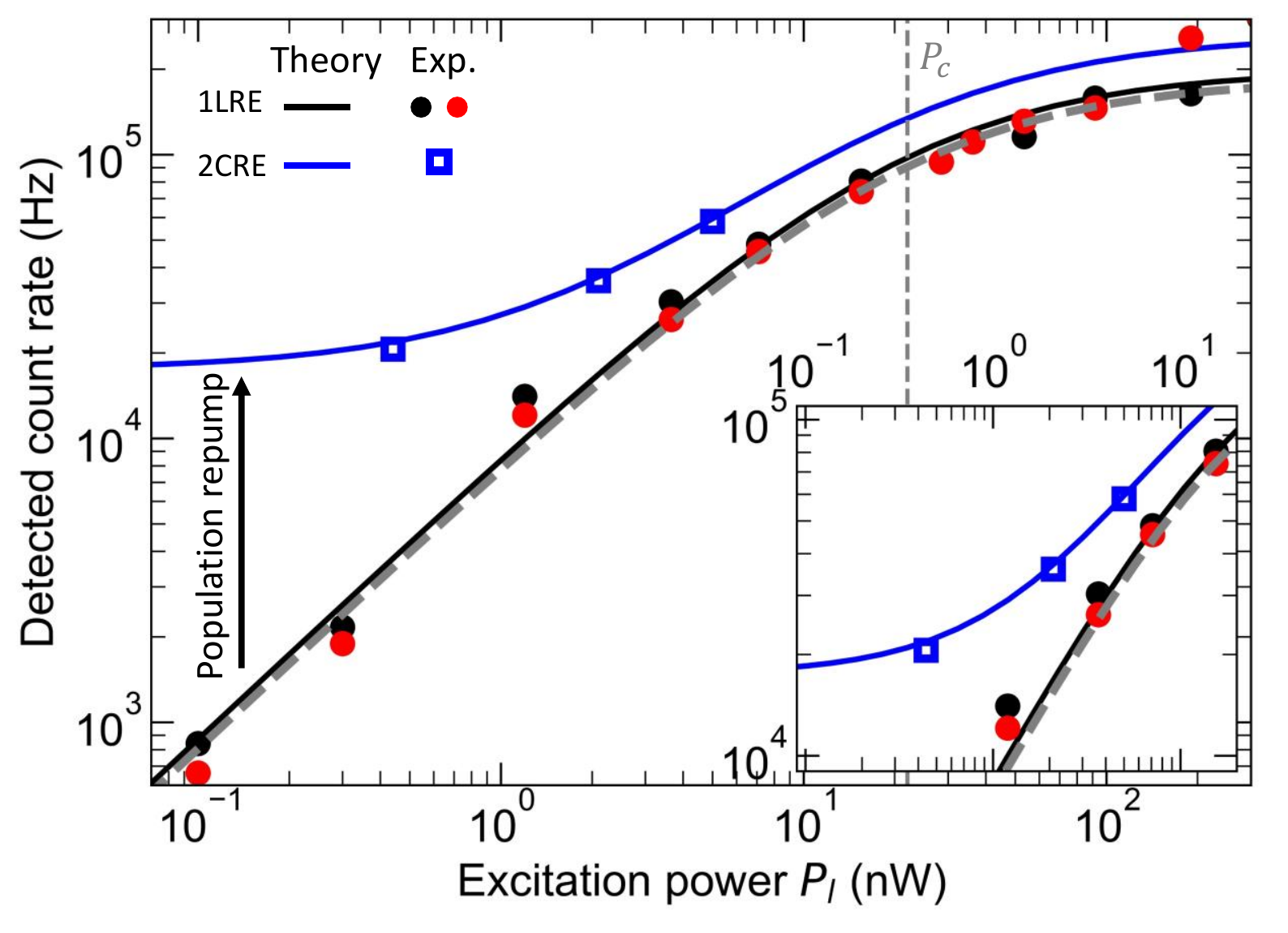}
\par\end{centering}
\caption{Power dependency of the trion resonant fluorescence under different
excitation schemes, comparing experimental photon count rates (symbols)
to our model (lines) with $\gamma_{\textrm{e}}=2.5\,\textrm{MHz}$:
Only pump laser for both trion transition (1LRE, black and red) and
with repump laser (2CRE, blue). The gray dashed lines indicate the
standard two-level system saturation behaviour. \label{Fig5:Power_dep_detected_rates}}
\end{figure}

Finally, we test our model against a series of excitation-power-dependent
experiments shown in Fig. \ref{Fig5:Power_dep_detected_rates}. Both
observed trion transitions under 1LRE (black and red symbols corresponding
to lines in Fig. \ref{Fig2:Voltage_scan_0T_vs_500mT}(d)) show saturation
with power described by $180\,\mathrm{kHz}/(1+P_{\textrm{c}}/P)$
\cite{loudon_quantum_1973,Snijders2018} with a reasonable saturation
power of $P_{\textrm{c}}=22\pm2\,\mathrm{nW}$, in agreement to our
model.

In contrast to these single-frequency measurements, the 2CRE scheme
shown by the blue symbols in Fig. \ref{Fig5:Power_dep_detected_rates}
shows clear signs of spin repumping: Due to the continuous repumping
of the spin population of both ground states with the two lasers (at
a constant $P_{r}=2.0\,\mathrm{nW}$), we control the individual steady-state
spin populations by altering the relative power of the lasers. Because
higher repumping power leads to stronger repumping and thus to higher
excited-state occupation, we experimentally observe increased photon
rates, following our model predictions. This increase varies with
relative powers between pump and repump laser beam from a factor higher
than 10 at $P_{l}=0.44\,\textrm{nW}$ to factor $1.3$ above $P_{\textrm{c}}$.

\section{Conclusions}

We developed a compact cryogenic SmCo permanent magnet assembly delivering
an in-plane magnetic field of $475\,\mathrm{mT}$. In contrast to
superconducting solenoids, this solution does not need any active
control and works from cryogenic to ambient temperatures. Therefore,
we believe it could become a preferable, economical, and scalable
architecture for spin-photon interfaces where the magnetic field is
used in “set-and-forget” mode.

Using this magnetic assembly in Voigt geometry, we have shown Zeeman
splitting and spin addressability of the electron and trion states
of a negatively charged quantum dot embedded in a birefringent optical
microcavity. We demonstrate spin-state manipulation using continuous-wave
resonant two-laser spectroscopy, which in combination with a high-extinction
ratio cross-polarization technique enables background-free single-photon
readout. This two-laser excitation scheme, similar to earlier schemes
\cite{Xu2007,Xu2008,Kroner2008PRB} without a cavity, will allow for
spin-state initialization and manipulation.
\begin{acknowledgments}
We acknowledge for funding from the European Union’s Horizon 2020
research and innovation programme under grant agreement No. 862035
(QLUSTER), from FOM-NWO (08QIP6-2), from NWO/OCW as part of the Frontiers
of Nanoscience program, from the Quantum Software Consortium, QuantumDelta,
and from the National Science Foundation (NSF) (0901886, 0960331).
\end{acknowledgments}


\pagebreak\renewcommand{\thefigure}{A\arabic{figure}}\setcounter{figure}{0}\renewcommand{\theequation}{A\arabic{equation}}\setcounter{equation}{0}
\appendix

\section{Permanent magnet assembly simulations\label{sec:Room-temperature-permanent}}

\begin{figure}
\begin{centering}
\includegraphics[width=1\columnwidth]{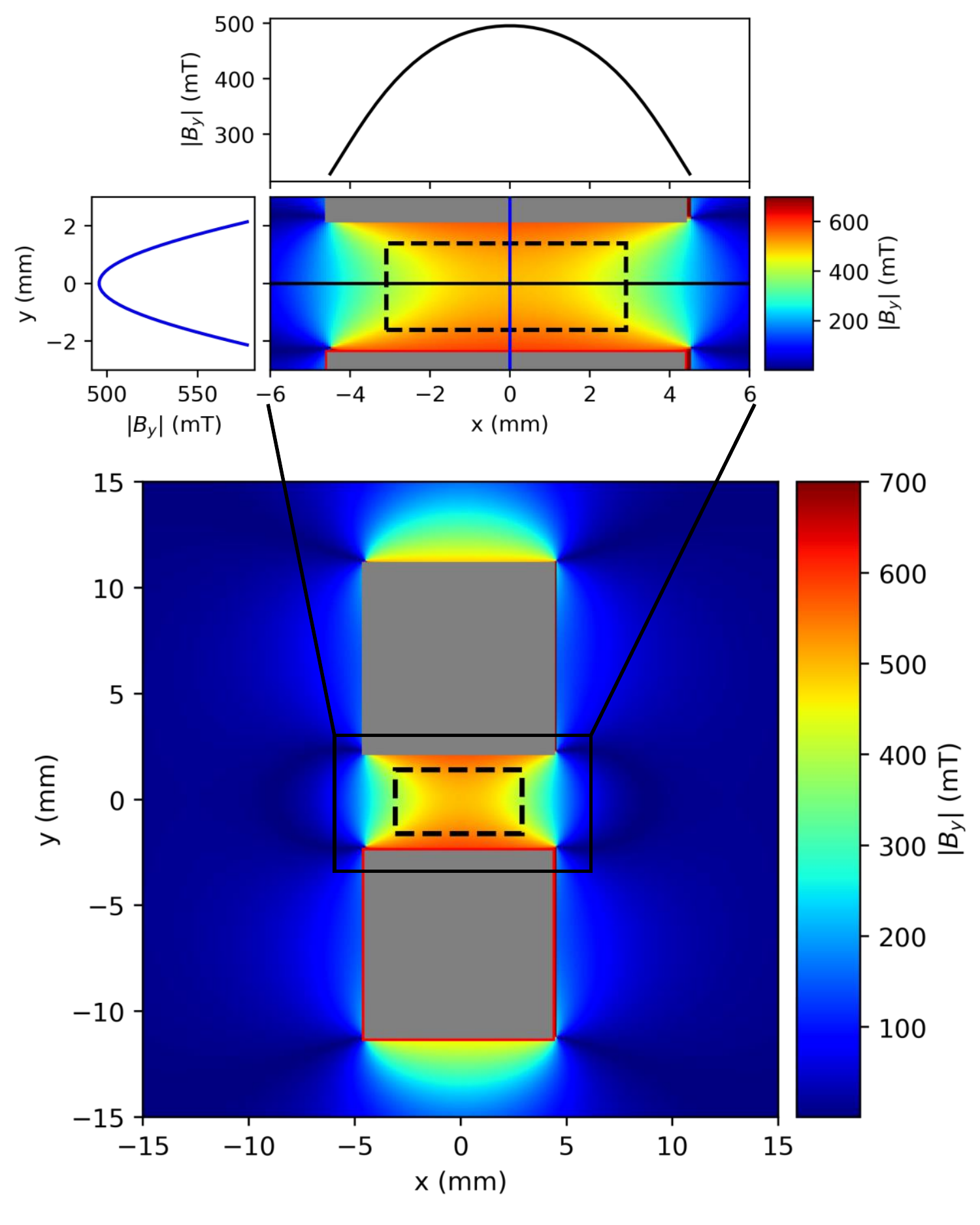}
\par\end{centering}
\caption{Magnetic field simulation of the magnetic assembly with a $4.5\,\mathrm{mm}$
air gap at room temperature. The magnitude of the magnetic field strength
$|B_{y}|$ along an $xy$-cross-section of the assembly (grey regions),
with the location of the sample taken to be the origin (outlined by
the dashed line). (Inset) Zoom-in $|B_{y}|$ to the sample region
with cross-sections along the $x$ (top) and $y$ (left) direction
through the center of the sample. \label{Fig1:MagnetSimu}}
\end{figure}
The magnetic assembly was simulated using Magpylib – a Python package
for magnetic field computation \cite{magpylib2020}. Given the large
and thermally stable coercivity of SmCo magnets at cryogenic temperatures
\cite{Liu2019,Wang2021}, we model the permanent magnets as $9\times9\:\mathrm{mm}$
large rods insensitive to any external magnetic field with a residual
magnetization of $1.03\,\mathrm{T}$. The copper housing of the magnet
was not included in the simulations, because copper is a weak magnetic
metal with low magnetic susceptibility \cite{Bowers1956}. The room
temperature simulation of our magnetic mount with a $4.5\,\mathrm{mm}$
air gap between the magnetic rods is presented in Fig. \ref{Fig1:MagnetSimu}.
From the simulation, we see that the assembly produces a strong magnetic
field (beyond $500\,\mathrm{mT}$) confined between the poles of the
magnets. Due to the simple assembly design, the magnetic field is
inhomogeneous over the entire sample footprint of several square millimeters.
However, over the few nanometer-size quantum dot used later in the
experiments, the magnetic field can be assumed homogenous. In our
experiments, we use QD close to the coordinate origin in Fig. \ref{Fig1:MagnetSimu},
where the external magnetic field reaches a strength of $500\,\mathrm{mT}$.

\begin{figure}
\begin{centering}
\includegraphics[width=1\columnwidth]{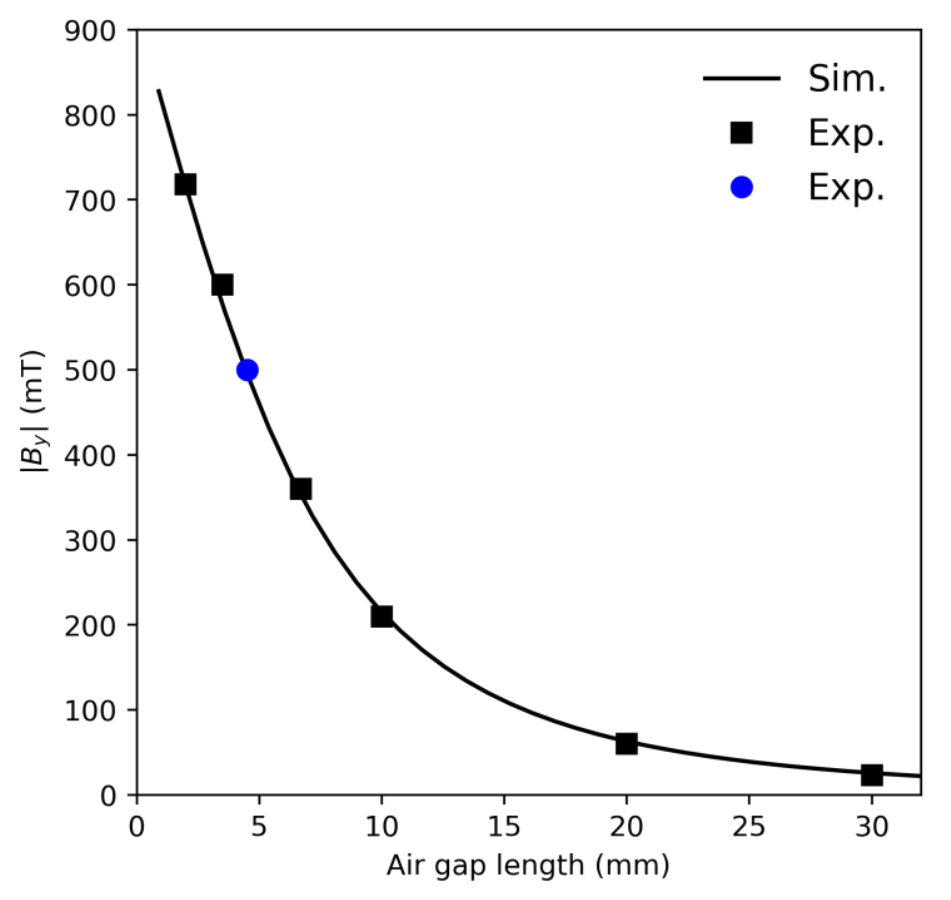}
\par\end{centering}
\caption{Air gap length dependence of magnetic field strength $|B_{y}|$. The
experimental Hall probe data points taken before (black) and after
(blue) fixing SmCo magnets into copper mount are compared to at the
sample center simulated $|B_{y}|$ (curve). \label{Fig1:Magnet_airgap}}
\end{figure}
The external field can be tuned by the air gap length, as shown in
Fig. \ref{Fig1:Magnet_airgap}. First, before mounting the magnets
into the copper housing, we fix a Hall probe to the center of the
air gap and vary the gap length. The measured field strengths excellently
agree with our simulations for various air gaps. Finally, the rods
are glued at the distance of $4.5\,\mathrm{mm}$ into the copper housing,
and a field of $500\,\mathrm{mT}$ in the air gap center is confirmed
by Hall probe measurements.

\renewcommand{\thefigure}{B\arabic{figure}}\setcounter{figure}{0}\renewcommand{\theequation}{B\arabic{equation}}\setcounter{equation}{0}

\section{Experimental setup and characterization \label{subsec:Setup-transmission-measurement}}

\begin{figure}
\begin{centering}
\includegraphics[width=1\columnwidth]{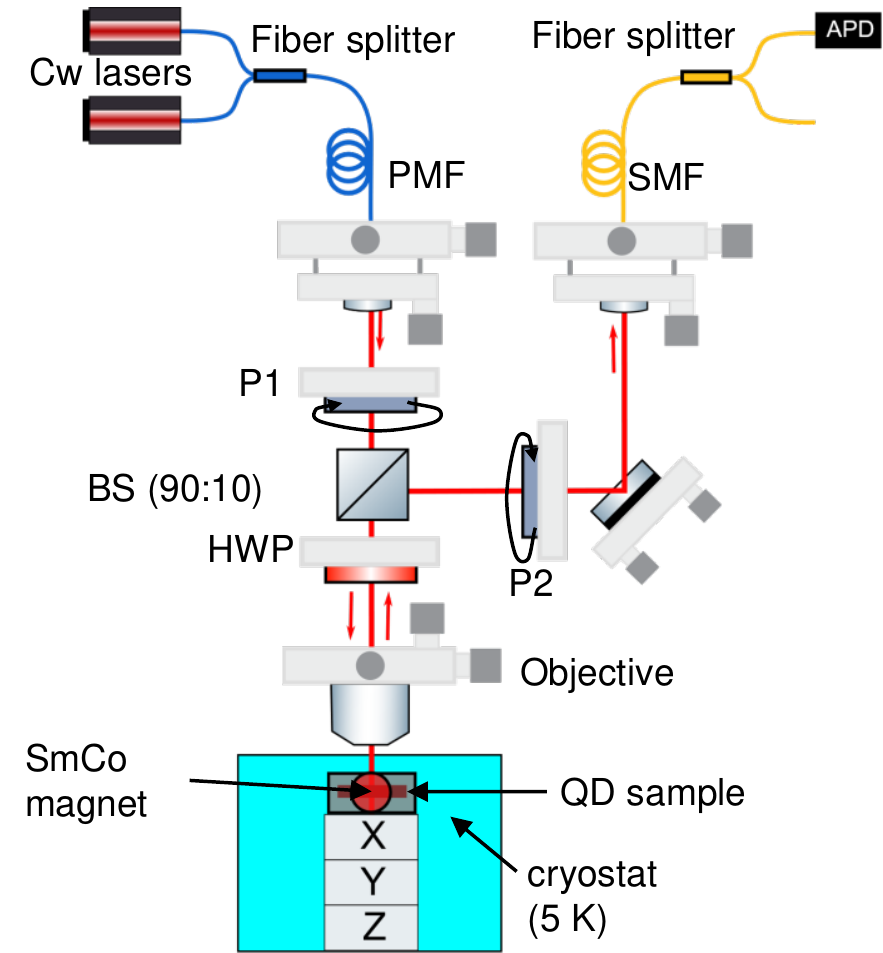}
\par\end{centering}
\caption{Experimental setup. \label{Fig:Setup_transmission}}
\end{figure}

For all our resonant fluorescence experiments, we use a confocal microscope
\cite{Steindl2023_PER} sketched in Fig. \ref{Fig:Setup_transmission}.
Here, two continuous wave narrow-linewidth ($200\,\textrm{kHz}$)
scanning lasers are fiber coupled to polarization maintaining fibers
(PMF), combined on polarization maintaining fiber splitter, and launched
into the vertical confocal microscope. The laser light is directed
on a free space non-polarizing beamsplitter (BS, splitting ratio 90:10
with transmission $\eta_{\textrm{BS,T}}=0.1$) and focused through
two silica windows into closed-cycle cryostat with a long-distance
working distance ambient-temperature objective with a total transmission
of $\eta_{\textrm{obj}}=0.62$. The excitation polarization is controlled
and aligned along the $V$ cavity mode with a Glan-Thompson polarizer
(P1) and zero-order half-waveplate (HWP; @$935\,\textrm{nm}$, quartz,
transmission > 0.99), both mounted in finely tunable motorized rotation
stages with a resolution of $10\,\textrm{mdeg}$. The last transmission
we need to consider is the fraction of the light transmitted through
the top mirror of the cavity. We estimated this transmission from
the distributed Bragg reflector design as $T_{\textrm{cav}}=3.4\times10^{-4}$
\cite{Bakker2015}. Then, the measured $21\,\mathrm{nW}$ optical
excitation power in front of the BS corresponds to an excitation power
of $0.44\,\mathrm{pW}$ at the location of the QD.

The photons emitted by the quantum dot and the reflected laser are
reflected at the BS with reflectivity $\eta_{\textrm{BS,R}}=0.9$.
The QD resonant fluorescence is separated from the excitation laser
using a cross-polarization scheme, where the excitation laser is rejected
by a factor $4\times10^{6}$ by using a nanoparticle polarizer (P2;
transmission $\eta_{\textrm{P}}=0.9$) in a motorized rotation stage
with $1\,\textrm{mdeg}$ resolution. Due to the alignment of the magnetic
field, we assume that the linearly polarized trion transitions are
perfectly aligned with the cavity polarization axes. Thus, the emission
from the two transitions with the same polarization as the excitation
laser is perfectly filtered out, while emission from the two orthogonal
transitions is fully transmitted. The separated emission from the
QD is then fiber coupled in a single-mode fiber (SMF; coupling efficiency
0.85, including collimation-lens transmission) and sent through a
fiber splitter on a single-photon detector (APD; $\eta=0.25$). Due
to loss in the fiber-splitter, the total free space-to-detector collection
efficiency is 0.32. The total transmission through the optical detection
system is $\eta_{\textrm{det}}=0.32\eta_{\textrm{obj}}\eta_{\textrm{BS,R}}\eta_{\textrm{P}}\eta=0.04$.

\renewcommand{\thefigure}{C\arabic{figure}}\setcounter{figure}{0}\renewcommand{\theequation}{C\arabic{equation}}\setcounter{equation}{0}

\section{Single-laser resonance fluorescence \label{subsec:Single-laser-resonant}}

The in-plane magnetic field of $475\,\mathrm{mT}$ splits the studied
trion transition via the Zeeman effect into two pairs of linearly
polarized emission lines with mutually orthogonal polarization. We
observe a splitting of $3.4\pm0.1\,\mathrm{GHz}$ and $1.8\pm0.1\,\mathrm{GHz}$
between $V$-polarized and $H$-polarized transitions, respectively,
corresponding to electron and hole g-factors of $|g_{\textrm{e}}|=0.39$
and $|g_{\textrm{h}}|=0.12$. 

\begin{figure}
\begin{centering}
\includegraphics[width=1\columnwidth]{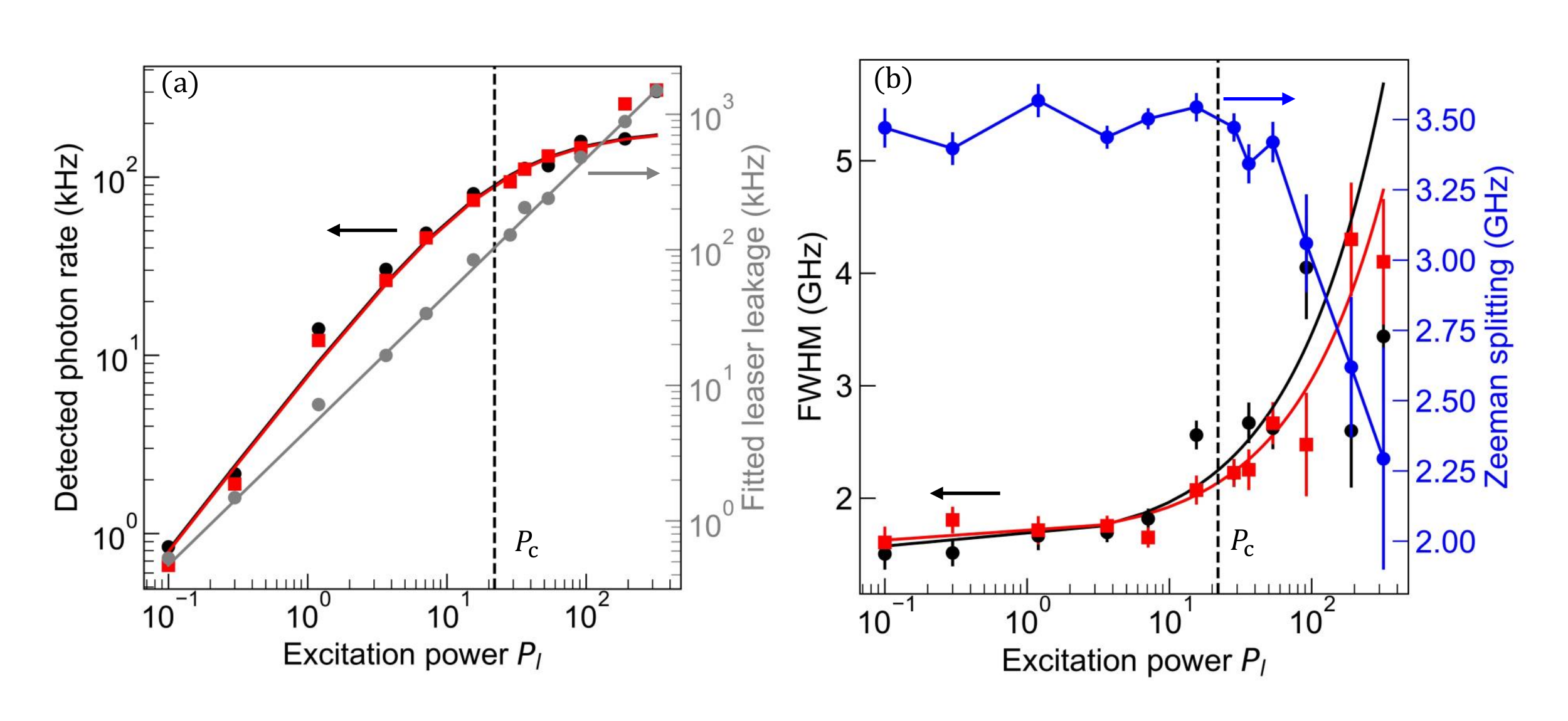}
\par\end{centering}
\caption{Single-laser power dependent characterization. Panel (a) shows how
the detected rate of photons from the two trion transitions (black,
red) and laser leakage (grey) depends on laser power. Panel (b) shows
the power dependence of the full width at half maximum of these two
Lorentzian transitions (black, red), and the Zeeman splitting (blue).
Error bars show the statistical error of the fit parameters and solid
lines show the model fits. \label{Fig:1laser_RF_power}}
\end{figure}

In the main text, we mainly focus on $V$-polarized resonant excitation
with varied excitation power. We observe a constant Zeeman splitting
over a bias range of more than $200\,\mathrm{mV}$, therefore we characterize
the excitation power properties only for a bias voltage of $1.49\:\mathrm{V}$,
a voltage where the transitions are in resonance with the $H$-polarized
cavity mode. The pair of trion emission lines is detected in cross-polarization
under $V$-polarized excitation of varied optical power $P_{l}$ over
three orders in magnitude. We fit the measured resonance fluorescence
spectrum with double Lorentzian function with a constant term characterizing
an excitation laser leakage due to finite cross-polarization extinction
ratio, and present the power dependency of the individual fit parameters
in Fig. \ref{Fig:1laser_RF_power}. We observe near-identical behavior
for the emission lines in both photon rate and line broadening. Figure
\ref{Fig:1laser_RF_power}(a) shows the detected photon rate, which
is well fit by $180\,\mathrm{kHz}/(1+P_{\textrm{c}}/P_{l})$ \cite{loudon_quantum_1973},
characterizing the two-level system saturation at power $P_{\textrm{c}}=22\pm2\,\mathrm{nW}$.
Similarly to our previous work \cite{Snijders2018}, we observe a
power-linear background (gray), most likely due to imperfect polarization
extinction. In Fig \ref{Fig:1laser_RF_power}(b), we analyze excitation-power
induced linewidth (FWHM) broadening. The experimental data show a
linewidth of $\Gamma=1.55\pm0.1\,\mathrm{GHz}$ at low excitation
power, with a significant broadening above $P_{\textrm{c}}$. This
broadening is well described with a simple power-law model $\Gamma+\beta P_{l}^{2/3}$
\cite{Sauncy1999,Huang2009}, using a parameter $\beta=(77\pm10)\times10^{3}\,\mathrm{GHz^{3/2}W^{-3/2}}$,
and can be caused by an increase in the dephasing rate induced by
nuclei polarization. The variation of the polarization of the nuclear-spin
bath will also affect the eigenenergies, leading to significant changes
in Zeeman splitting, as observed in Fig. \ref{Fig:1laser_RF_power}(b).

\renewcommand{\thefigure}{D\arabic{figure}}\setcounter{figure}{0}\renewcommand{\theequation}{D\arabic{equation}}\setcounter{equation}{0}

\section{Rate-equation model of resonant two-color spectroscopy of a negatively
charged exciton\label{subsec:Semi-classical-model-of}}

\begin{figure}
\centering{}\includegraphics[width=1\columnwidth]{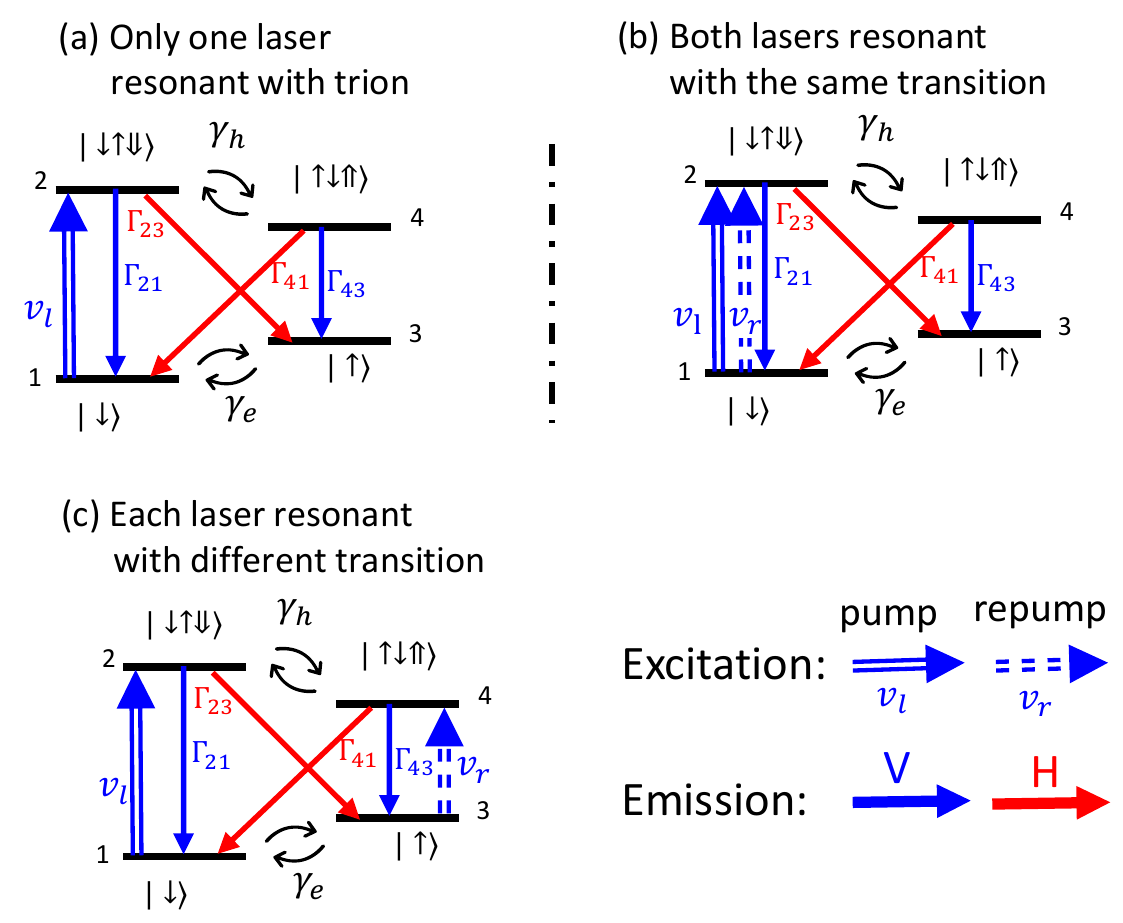}\caption{Resonant excitation schemes of trion: (a) a single excitation laser
is resonant with a trion transition, (b) two lasers of identical polarization
are resonant with the same transition, (c) two lasers are resonant
with distinct trion transitions. \label{Fig:Double_resonant_excitation_scheme}}
\end{figure}

In this section, we describe our theoretical model used for comparison
and understanding of the two-color resonance fluorescence experiments.
Limiting the description to continuous-wave (cw) resonant excitation
of the trion states in Voigt configuration, we model the trion energy
levels as two coupled $\Lambda$ systems. Figure \ref{Fig:Double_resonant_excitation_scheme}
shows a sketch of the interaction, where in total 4 optical transitions
in a linear basis are possible: two emitting $V$-polarized (blue)
and two emitting $H$-polarized photons (red), respectively. In addition
to these optical transitions, there are also two spin-flip transitions,
one for electron spin $\gamma_{\textrm{e}}$ and one for hole spin
$\gamma_{\textrm{h}}$.

In the experiment, a strong laser power was used, therefore we can
neglect quantization of the excitation light together with stimulated
emission, but it was kept at least factor 3 below saturation intensity
$P_{\textrm{c}}$. Within this limit, we can separate the problem
into two steps: (i) setting up and solving rate equations characteristic
to individual energy level configurations, and (ii) expression of
emitted photon rates based on state populations found from (i).

\subsubsection{Spin population rate equations}

The modeling of the scanning two-color resonant excitation of two
coupled $\Lambda$ systems can be split into three scenarios, depicted
in Fig. \ref{Fig:Double_resonant_excitation_scheme}, distinguished
by which transitions are addressed with the excitation lasers. Additionally,
in correspondence to our experiment, the model is developed only for
$V$-polarization, reducing the complexity.

We start with a situation when only a single laser is resonant with
the trion energy levels, as depicted in Fig. \ref{Fig:Double_resonant_excitation_scheme}(a).
For simplicity, we discuss here only resonant excitation of the $|\downarrow\rangle\rightarrow|\downarrow\uparrow\Downarrow\rangle$
transition, $|\uparrow\rangle\rightarrow|\uparrow\downarrow\Uparrow\rangle$
can be derived easily. Here, the population is brought from the ground
state $|\downarrow\rangle$ to the excited state with a single resonant
laser of excitation rate $v_{l}$. The excited state relaxes back
to the $|\downarrow\rangle$ or $|\uparrow\rangle$ spin state by
spontaneous emission of a $V$ (emission rate $\Gamma_{21}$) and
$H$ (emission rate $\Gamma_{23}$) polarized single photon, or via
a hole spin-flip transition to $|\uparrow\downarrow\Uparrow\rangle$.
We use the steady-state condition to solve the trion-state population
described by the interaction matrix
\[
M_{\mathrm{pump}}=\left(\begin{array}{cccc}
-(v_{l}+\gamma_{\textrm{e}}) & \Gamma_{21} & \gamma_{\textrm{e}} & \Gamma_{41}\\
v_{l} & -(\Gamma_{2}+\gamma_{\textrm{h}}) & 0 & \gamma_{\textrm{h}}\\
\gamma_{\textrm{e}} & \Gamma_{23} & -\gamma_{\textrm{e}} & \Gamma_{43}\\
0 & \gamma_{\textrm{h}} & 0 & -(\Gamma_{4}+\gamma_{\textrm{h}})
\end{array}\right)
\]
and analytically find the state population $P_{|x\rangle}$ for each
trion state involved in trion dynamics:
\begin{align}
P_{|\downarrow\rangle} & =\frac{\gamma_{\textrm{e}}\Gamma_{2}\Gamma_{4}+\gamma_{\textrm{e}}\gamma_{\textrm{h}}(\Gamma_{2}+\Gamma_{4})}{\alpha(v_{l})},\nonumber \\
P_{|\downarrow\uparrow\Downarrow\rangle} & =\frac{\gamma_{\textrm{e}}(\Gamma_{4}+\gamma_{\textrm{h}})}{\alpha(v_{l})}v_{l},\nonumber \\
P_{\left|\uparrow\right\rangle } & =P_{|\downarrow\rangle}+\frac{\Gamma_{23}}{\gamma_{\textrm{e}}}P_{|\downarrow\uparrow\Downarrow\rangle}+\frac{\Gamma_{43}\gamma_{\textrm{h}}}{\alpha(v_{l})}v_{l},\nonumber \\
P_{|\uparrow\downarrow\Uparrow\rangle} & =\frac{\gamma_{\textrm{e}}\gamma_{\textrm{h}}}{\alpha(v_{l})}v_{l}.\label{eq:Population_1laser}
\end{align}
Here, we use the total emission rates $\Gamma_{2}=\Gamma_{21}+\Gamma_{23}$
and $\Gamma_{4}=\Gamma_{41}+\Gamma_{43}$ from excited states $|\downarrow\uparrow\Downarrow\rangle$
and $|\uparrow\downarrow\Uparrow\rangle$, together with $\alpha(x)=[\Gamma_{43}\gamma_{\textrm{h}}+\Gamma_{23}(\Gamma_{4}+\gamma_{\textrm{h}})+\gamma_{\textrm{e}}\Gamma_{4}(2\Gamma_{2}+1)+2\gamma_{\textrm{e}}\gamma_{\textrm{h}}(\Gamma_{4}+\Gamma_{2}+1)]x$
to simplify the notation.

Now, we focus on two-laser excitation of the same transition, Fig.
\ref{Fig:Double_resonant_excitation_scheme}(b). Here, the two lasers
have identical frequency and polarization and differ only in optical
power, therefore we can model them as a single laser of optical power
corresponding to $v_{l}+v_{r}$, where $v_{l}$ and $v_{r}$ are the
excitation rates of pump and repump lasers. Then the state occupations
have again form of eq.$\,$(\ref{eq:Population_1laser}), with the
only change in the excitation rate $v_{l}\rightarrow v_{l}+v_{r}$.

For the two-color excitation scheme, where each laser is in resonance
with a distinct trion transition, as sketched in Fig. \ref{Fig:Double_resonant_excitation_scheme}(c).
Because the interaction matrix 
\[
M_{\mathrm{pump\&repump}}=M_{\mathrm{pump}}+\left(\begin{array}{cccc}
0 & 0 & 0 & 0\\
0 & 0 & 0 & 0\\
0 & 0 & -v_{r} & 0\\
0 & 0 & v_{r} & 0
\end{array}\right)
\]

does not have a steady-state analytical solution, we obtain the state
occupations $P_{|x\rangle}$ numerically.

\subsubsection{2D two-color resonant excitation model}

Now we formulate a simple model interconnecting our two-color resonance
fluorescence experiment with the steady-state trion occupations derived
from the system rate equations. First, we assume that the emitted
resonance fluorescence rate is proportional to excited state occupations
and the radiative transition rates as
\[
I=(f_{V}\Gamma_{21}+f_{H}\Gamma_{23})P_{|\downarrow\uparrow\Downarrow\rangle}+(f_{V}\Gamma_{43}+f_{H}\Gamma_{41})P_{|\uparrow\downarrow\Uparrow\rangle.}
\]
Here we use a parameter $f_{x}$ where the subscript indicates the
emitted photon polarization allowing later implementation of the cross-polarization
scheme by setting $f_{V}=0$ and $f_{H}=1$. We assume that the pair
of the observed emission lines is resonantly excited with a laser
of frequency $f_{1}^{\textrm{QD}}$ and $f_{2}^{\textrm{QD}}$, and
each of the lines has a Lorentzian shape characterized by an identical
full width at half maximum $\Gamma$, in agreement with our previous
experiments in Sec. \ref{subsec:Single-laser-resonant}. Using $\Gamma$,
$f_{1}^{\textrm{QD}}$, and $f_{2}^{\textrm{QD}}$ from single-laser
resonance fluorescence experiments, we model the emission as 2D Lorentzian
functions $\mathrm{L}(x,x_{0},y,y_{0},\Gamma)=\frac{2}{\pi\Gamma}[(x-x_{0})^{2}+(y-y_{0})^{2}+(\Gamma/2)^{2}]^{-1}$
multiplied with photon rate $I$ calculated from the rate equations.
As discussed above, the rate equations and thus also the state occupations
and $I$ varies with specific resonant excitation configuration. The
different conditions we label by $I^{(i,j)}$, where $i,j$ indicate
with which transitions the laser are resonant with ($f_{x}^{\textrm{QD}}$)
or 0 if the laser is not resonant with any trion transition. The final
two-laser model is given by
\begin{align}
I_{\mathrm{total}=} & \sum_{i\in\{1,2\}}I^{(i,0)}\mathrm{L}(f_{r},f_{i}^{\textrm{QD}},0,0,\Gamma)\nonumber \\
 & +\sum_{i\in\{1,2\}}I^{(0,i)}\mathrm{L}(0,0,f_{l},f_{i}^{\textrm{QD}},\Gamma)\nonumber \\
 & -\sum_{i,j\in\{1,2\}}(I^{(i,0)}+I^{(0,j)})\mathrm{L}(f_{r},f_{i}^{\textrm{QD}},f_{l},f_{j}^{\textrm{QD}},\Gamma)\label{eq:2laser_RFmodel}\\
 & +\sum_{i\in\{1,2\}}I^{(i,i)}\mathrm{L}(f_{r},f_{i}^{\textrm{QD}},f_{l},f_{i}^{\textrm{QD}},\Gamma)\nonumber \\
 & +\sum_{i,j\in\{1,2\},i\neq j}I^{(i,j)}\mathrm{L}(f_{r},f_{i}^{\textrm{QD}},f_{l},f_{j}^{\textrm{QD}},\Gamma).\nonumber 
\end{align}
Here, the first two terms describe emission under single-laser excitation
with separate lasers, the third term removes contributions of the
individual lasers that would be counted twice otherwise. The fourth
term accounts for emission by simultaneous two-resonant laser excitation
of the identical transition, and the fifth term for concurrent two-color
excitation of two distinct transitions.

\subsubsection{Estimate of excitation and detection rates}

To connect the theoretical model with our experiment, we need to estimate
the trion driving power from optical power measured in the setup.
It requires conversion of an optical power measured with a power meter
to the individual laser excitation rates $v_{l}$ and $v_{r}$. First,
we determine the setup throughput as described in Sec. \ref{subsec:Setup-transmission-measurement}.
As an example, the optical power is $P=21\,\mathrm{nW}$, measured
in our setup in front of BS. Using the measured transmission of the
excitation path of our setup ($\eta_{\textrm{ex}}=T_{\textrm{cav}}\eta_{\textrm{obj}}\eta_{\textrm{BS,T}}$),
and assuming unity QD quantum efficiency, we estimate that the QD
is excited with an optical power corresponding to $\eta_{\textrm{ex}}P=0.44\,\mathrm{pW}$.
The excitation rate is then calculated from this power by multiplication
with an experimentally determined conversion factor between power-meter
readings and the single-photon rate measured with a single-photon
detector.

Similarly, we correct the theoretical emission for the detection system
optical throughput simply by its multiplication with experimentally
determined throughput $\eta_{\textrm{det}}$.

\subsubsection{Model rates estimation}

We start the discussion with the radiative rates of our trion-cavity
system. An isolated trion in Voigt geometry typically has all four
radiative transitions of an identical rate around $\Gamma_{0}=1\,\mathrm{GHz}$
\cite{Bayer2002}. The situation is different if a trion is coupled
into a linearly polarized cavity mode leading to Purcell enhancement.
Neglecting pure dephasing, we estimate the cavity-enhanced rates from
the QD emission line width $\Gamma$ of $1.5\:\mathrm{GHz}$, giving
$\Gamma_{21}=\Gamma_{43}=3\,\mathrm{GHz}$. Assuming that the second
pair of rates correspond to an isolated trion, we estimate these rates
to be $\Gamma_{23}=\Gamma_{41}=1\,\mathrm{GHz}$. Since these rates
are very sensitive to the specific condition, we keep them as free
fit parameters for the two-color experiments show in the main text
and below.

Now we discuss how we estimate the electron and hole-spin flip rates
based on comparison of the power dependence of the single laser resonance
fluorescence with our model for single laser excitation, i.e., using
only the first term in eq. (\ref{eq:2laser_RFmodel}). We model the
trion level system with the radiative rates estimated above and vary
only $\gamma_{\textrm{e}}$ and $\gamma_{\textrm{h}}$.

\begin{figure}
\centering{}\includegraphics[width=1\columnwidth]{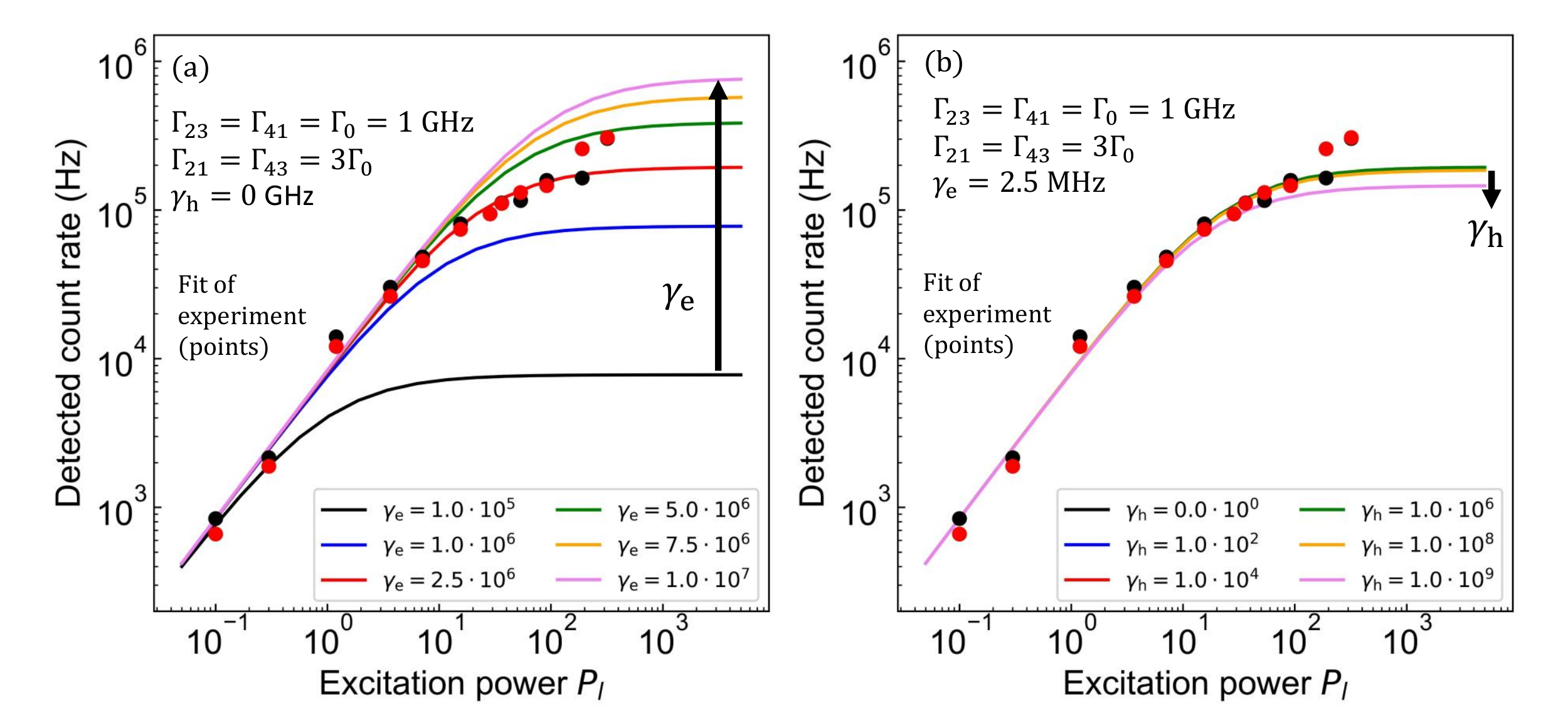}\caption{Spin-flip rate estimate from excitation power dependent trion emission
under single resonant excitation. Lorentzian fits of experimental
data (data points) are compared to our model with (a) varied $\gamma_{\textrm{e}}$
and fixed $\gamma_{\textrm{h}}=0$ to estimate $\gamma_{\textrm{e}}$
value, and in (b) with varied $\gamma_{\textrm{h}}$ and fixed $\gamma_{\textrm{e}}=2.5\,\mathrm{MHz}$.
\label{Fig:RF_1laser_spinflip_rates}}
\end{figure}

First, we estimate the electron spin-flip rate. Typically $\gamma_{\textrm{h}}\ll\gamma_{\textrm{e}}\ll\Gamma_{0}$
, therefore we can neglect the hole spin-flip transition and set in
our model $\gamma_{\textrm{h}}=0$. Then, we determine the most likely
value of $\gamma_{\textrm{e}}$ by comparing a simulated power dependence
of the detected rate for various $\gamma_{\textrm{e}}$ with the experimentally
observed rates, as shown in Fig. \ref{Fig:RF_1laser_spinflip_rates}(a).
In good agreement with $\gamma_{e}=1.2\,\mathrm{MHz}$ reported in
\cite{Kroner2008PRB}, we achieved the best agreement between the
model and experimental data for $\gamma_{\textrm{e}}=2.5\,\mathrm{MHz}$.
In Fig. \ref{Fig:RF_1laser_spinflip_rates}(b), we present a similar
simulation, now with fixed $\gamma_{\textrm{e}}=2.5\,\mathrm{MHz}$
and varied $\gamma_{\textrm{h}}$ to reveal the model dependency on
$\gamma_{\textrm{h}}$. In contrast to Fig. \ref{Fig:RF_1laser_spinflip_rates}(a),
we observe only weak dependence of the model on $\gamma_{\textrm{h}}$,
so we set in all our simulations for simplicity $\gamma_{\textrm{h}}=0$.
This agrees with the fact that we can only observe hole-spin flips
during the very short trion lifetime.

\subsubsection{Excitation-power dependent two-color resonant excitation}

Here we study the excitation power dependency of trion state occupations
under different resonant excitation schemes and compare the model
with the measured two-laser excitation resonance fluorescence. The
parameters to model the trion steady-state population are estimated
from a least-square fit (discussed below) of the experimental data
with an optical power of $P_{r}=2.0$ and $P_{l}=2.1\,\mathrm{nW}$,
and the best agreement is achieved for parameters $\Gamma_{21}=2.1\,\textrm{GHz},\Gamma_{43}=2.7\mathrm{\,GHz}$,
$\Gamma_{23}=\Gamma_{41}=0.8\,\mathrm{GHz}$, $\gamma_{\textrm{e}}=2.5\,\mathrm{MHz},$
and $\gamma_{\textrm{h}}=0\,\mathrm{Hz}$. 

\begin{figure}
\centering{}\includegraphics[width=1\columnwidth]{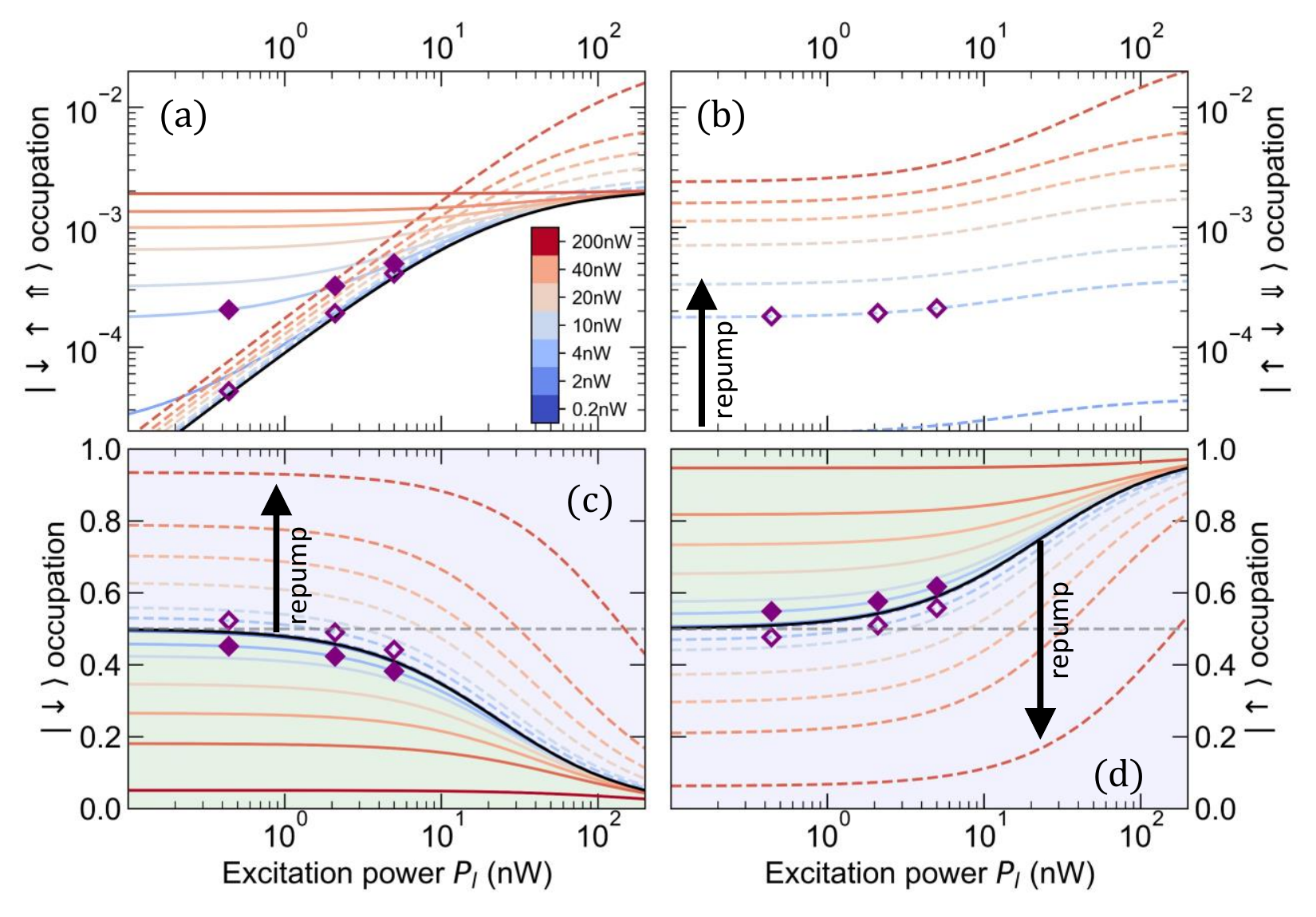}\caption{Simulation of the two-laser power dependency of trion state occupations.
Three different excitation schemes are compared: single laser excitation
(1LRE, black solid line), two-laser excitation of an identical line
(2LRE, colored solid lines), and two-color excitation of two distinct
transitions (2CRE, colored dashed lines). The power of the second
laser is encoded in the color scale. Two regions of ground state occupation
are highlighted (c,d): the orange region is accessible only with two-laser
excitation of an identical transition, and the blue one only if two
distinct transitions are resonantly excited. Purple diamonds (filled
if two lasers address the same transition, empty if different ones)
correspond to experimental conditions.\label{Fig:2LRF_StatesOccupation}}
\end{figure}

Assuming only minor power-induced rate changes for excitation below
$P_{\textrm{c}}$, in Fig. \ref{Fig:2LRF_StatesOccupation} we study
the state population of the individual trion levels as a function
of the optical power of both lasers. Under weak single-laser excitation
where the excitation rate is much slower than the radiative rate of
the transition, the radiative relaxation of the excited state into
both ground states is much faster than the excitation. Therefore the
dynamics is dominated by spontaneous emission. Since the radiative
rates of the transitions are approximately equal, we observe in (c)
and (d) an expected balanced ground state population of $\sim0.5$
\cite{Kroner2008PRB}. With increasing excitation power, repumping
of the $|\downarrow\rangle$ population into the excited state becomes
relevant, leading to a ground state population imbalance together
with a rise of the excited state population, which saturates at high
powers, see Fig. \ref{Fig:2LRF_StatesOccupation}(a). As discussed
also in the main text, the dynamics under two-laser excitation of
a single transition is equivalent to single-laser excitation with
higher excitation power. However, the dynamics changes under the two-color
excitation of two distinct transitions shown as dashed lines in Fig.
\ref{Fig:2LRF_StatesOccupation}, where the second laser repumps the
population from $\left|\uparrow\right\rangle $ to $|\uparrow\downarrow\Uparrow\rangle$.
As our model predicts, this repumping is higher with a stronger repumping
laser.

\begin{figure}
\centering{}\includegraphics[width=1\columnwidth]{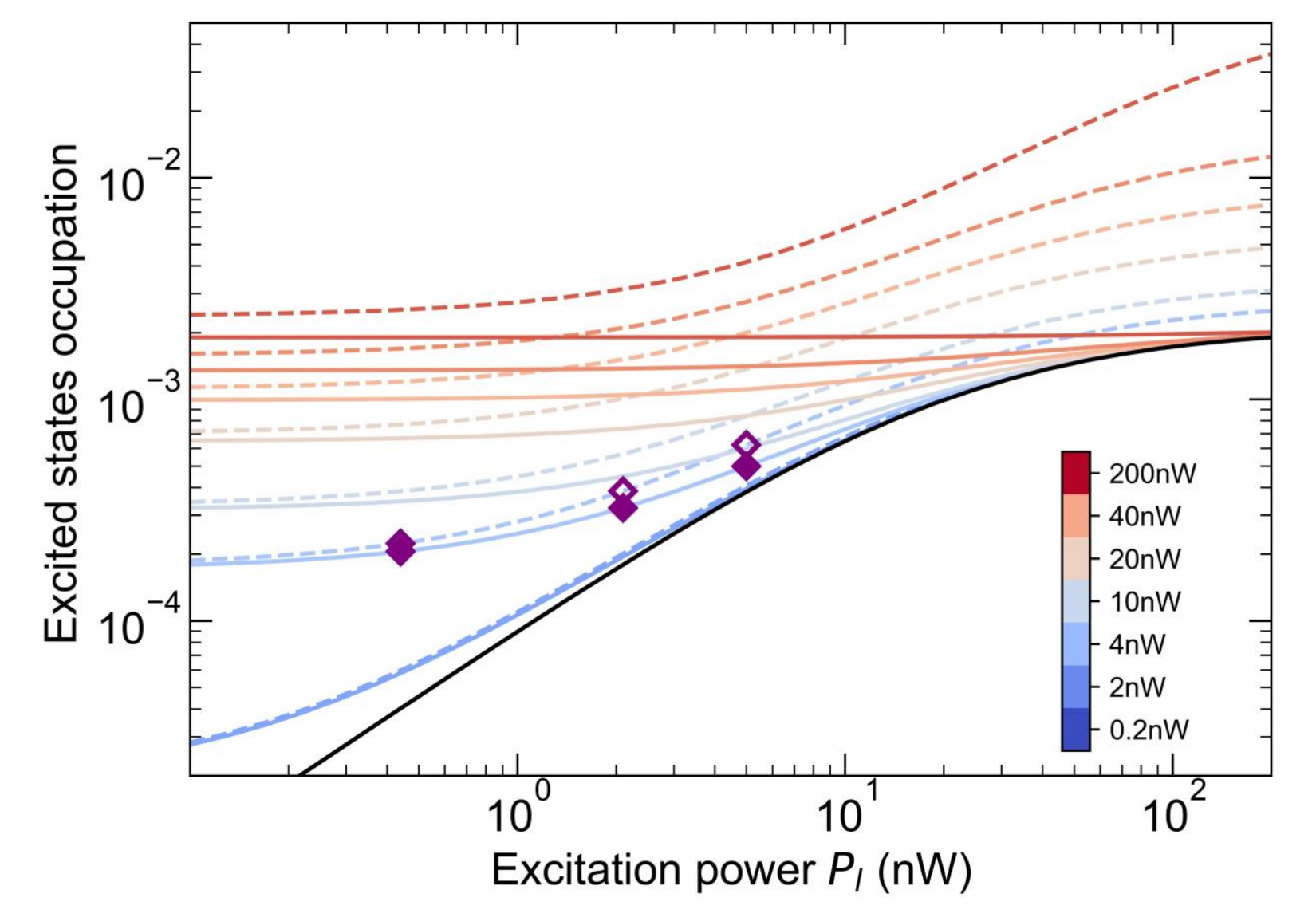}\caption{Simulated power dependency of the total excited states population.
Color and line style encoding is identical to that in Fig. \ref{Fig:2LRF_StatesOccupation}.
\label{Fig:ExcitedStatesOccupation}}
\end{figure}
Since emission is a measure of the population of the excited states,
we can gain insight about expected detected photon rates in different
resonant excitation schemes from the total occupation of excited states,
shown in Fig. \ref{Fig:ExcitedStatesOccupation}. Interestingly, the
total excited state population is always higher in the two-color excitation
scheme. That is because the second laser populates the second excited
state which was (due to negligible $\gamma_{\textrm{h}}$) not involved
in dynamics if only a single transition was resonantly pumped. This
shows that a repump laser can enhance the single-photon rate.

\begin{figure}
\centering{}\includegraphics[width=1\columnwidth]{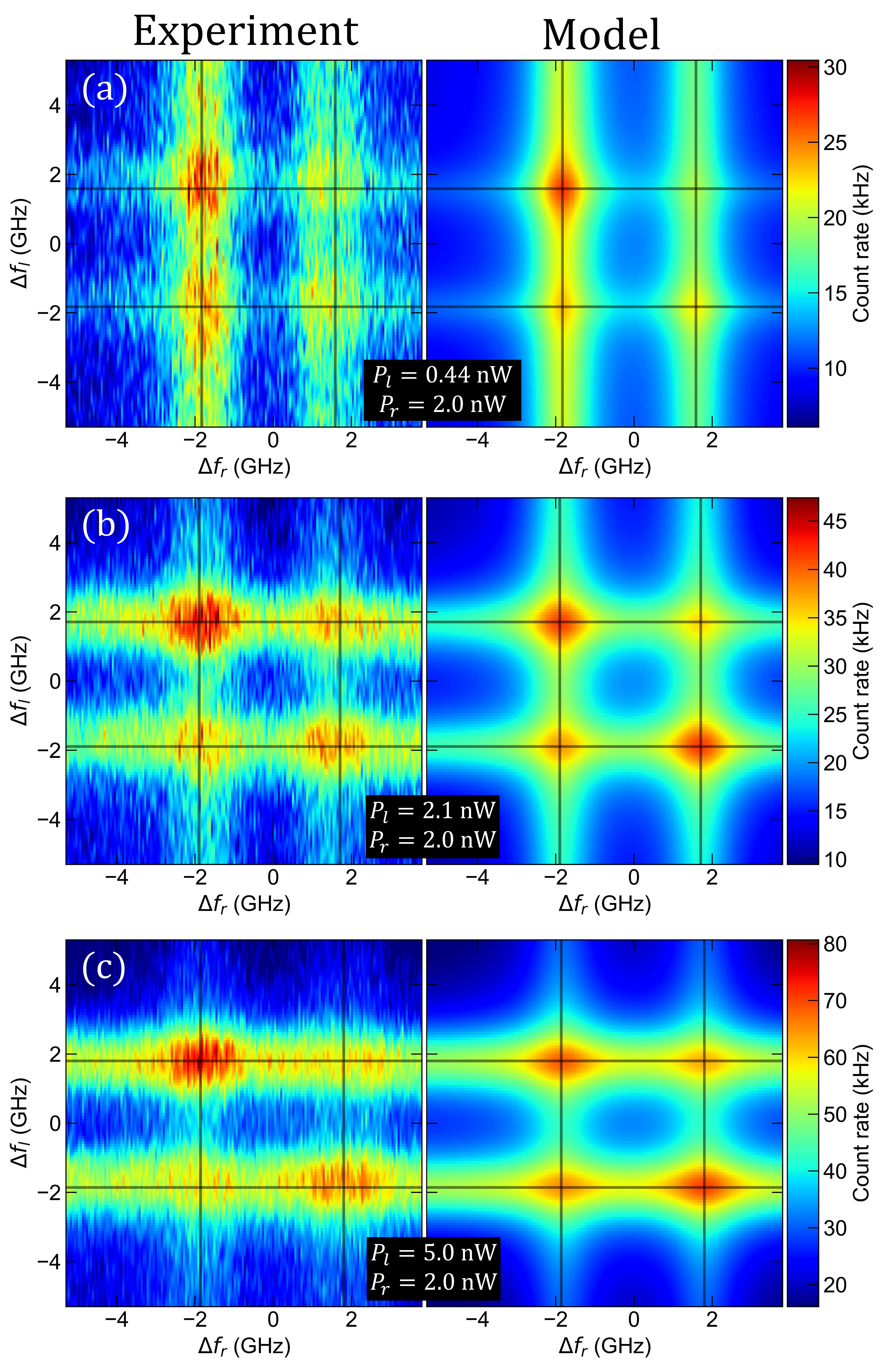}\caption{False-color plots of the resonant two-color laser scans and model
results. Experimental data is shown in the left column, the model
in the right column. The repump laser power is kept constant at $P_{r}=2.0\,\mathrm{nW}$,
while the pumplaser power $P_{l}$ is varied: (a) $P_{l}=0.44\,\mathrm{nW}$,
(b) $P_{l}=2.1\,\mathrm{nW}$, and (c) $P_{l}=5.0\,\mathrm{nW}$.
The transition frequencies of the QD are shown by solid black lines.
\label{Fig:2LRF_experiment_theory}}
\end{figure}
Finally, we discuss additional two-color experiments. In Fig. \ref{Fig:2LRF_experiment_theory},
we observe again a 'number sign' like structure, where horizontal
and vertical lines represent the trion transitions probed with probe
and pump laser, and at the intersections two-laser dynamics appears.
Again, spin repumping is clearly visible. We fit the model with an
extra term describing the background caused by imperfect cross-polarization
filtering on three sets of experimental data measured with fixed pump
laser power of $P_{r}=2.0\,\mathrm{nW}$ and we have varied optical
power of probe laser $P_{l}$. We use the following steps to achieve
the best agreement between the model and our experiment: First, we
fit the experiment using the initial estimate of radiative and spin-flip
rates, and QD linewidth and energies, as discussed above. We optimize
Zeeman splitting together with the linewidths, therefore we keep all
parameters (except $\gamma_{\textrm{h}}=0\,\mathrm{Hz}$) free. In
the next step, we fix the parameter describing the background, and
QD's $f_{1}^{\textrm{QD}}$, $f_{2}^{\textrm{QD}}$, $\Gamma$, and
optimize only radiative rates and $\gamma_{\textrm{e}}$. The best
fits of the model are compared to the experiment in Fig. \ref{Fig:2LRF_experiment_theory}.
Examination of the power-dependence of the parameters shows a power
broadening (FWHM) from $1.52\,\mathrm{GHz}$ to $1.89\,\mathrm{GHz}$
and a small increase of the Zeeman splitting from $3.40\,\mathrm{GHz}$
to $3.65\,\mathrm{GHz}$, as also shown in Fig. \ref{Fig:1laser_RF_power}.
The determined electron and hole spin-flip rates are $\gamma_{\textrm{e}}=2.5\:\mathrm{\textrm{MHz}}$
and $\gamma_{\textrm{h}}=0\,\mathrm{Hz}$; radiative rates are increasing
with excitation power and are $\Gamma_{21}=1.1-4.9\,\textrm{GHz},\Gamma_{43}=2.1-4.5\mathrm{\,GHz}$,
$\Gamma_{23}=\Gamma_{41}=0.7-0.8\,\mathrm{GHz}$, probably due to
the Rabi effect.
\end{document}